\documentclass{aa}

\usepackage{graphicx}

\usepackage{txfonts}

\usepackage[hidelinks,colorlinks=true,linkcolor=blue,allcolors=blue]{hyperref}

\begin{document}

   \title{Self-regulated galaxy evolution within a self-consistently varying galaxy-wide IMF}

   \author{Lukas Hof
          \inst{1}\fnmsep\inst{2}
          ,
          Pavel Kroupa
          \inst{1}\fnmsep\inst{3}
          ,
          Gerhard Hensler
          \inst{4}
          \and
          Jan Pflamm-Altenburg
          \inst{1}\\
          }

   \institute{Helmholtz-Institut für Strahlen- und Kernphysik, Universität Bonn, Nussallee 14-16, 53115 Bonn, Germany\\
              \email{lhof@astro.uni-bonn.de}
         \and
            Argelander-Institut für Astronomie, Universität Bonn, Auf dem Hügel 71, D-53121 Bonn, Germany
         \and
            Astronomical Institute, Faculty of Mathematics and Physics, Charles University, V Holešovičkách 2, CZ-180 00 Praha 8, Czech Republic
         \and
            Department of Astrophysics, University of Vienna, Türkenschanzstrasse 17, 1180 Vienna, Austria
             }

   \date{Received ...; accepted ...}

  \abstract
   {Semi-analytical evolution models of galaxies are a useful and computationally inexpensive tool for fast assessment of individual properties and their evolution. In this work, specifically the influence of a metallicity and star-formation rate (SFR) dependent galaxy-wide stellar initial mass function (IGIMF) on the self-regulation of star-formation in a galaxy is of interest.}
   {The impact of a variable IGIMF shall be investigated, especially its influence on the equilibrium SFR of the system, its gas fraction, the metallicity evolution and the gas depletion timescale, $\tau_{\rm gas}$, all in comparison with two non-varying galaxy-wide IMFs (gwIMF).}
   {A 2-phase 2-component model with gas and stellar components is integrated using the Cash-Karp method with adaptive step size. For comparison, the calculations are carried out using non-varying gwIMFs and a variable IGIMF, implemented using the \texttt{GalIMF} code which is based on the IGIMF theory. The input parameters are the effective radius, the final age and the accretion rate of the galaxy. Our highly simplified model does not represent a full physical galactic network, however it is suitable for the purpose of this study.}
   {All models -- both non-varying gwIMFs and the IGIMF -- reproduce reasonable gas fractions, gas depletion timescales and the main sequence of star-forming galaxies. However, only the IGIMF model accurately predicts the mass-metallicity relation and provides a more comprehensive description of quenched elliptical galaxies. For massive ellipticals all models suggest the need for an additional gas heating source to reach a quenched state. Using a different stellar yield table in the IGIMF model does not significantly affect the results. In all models, the galaxies evolve self-regulated, determined by the accretion rate. The self-regulated constancy of the SFR reflects the constant SFRs of nearby star-forming galaxies. The specific gas-accretion rate of all galaxies appears to be comparable to the Hubble constant. The inclusion of outflows improves the results for the canonical gwIMF model, but not significantly, while for the IGIMF model it has no significant impact.
   }
   {}

   \titlerunning{Self-regulated galaxy evolution}
   \authorrunning{L. Hof et al.}
   \maketitle
%

\section{Introduction}

    Semi-analytical chemical evolution models of galaxies are an important tool to achieve a better understanding of the physical evolution of galaxies. To understand such an evolution one needs a comprehensive yet not an overly complex description of the different phases and components of a galaxy as well as of the star-formation rate (SFR), which itself depends on the physical parameters.
    While self-consistent high-resolution hydrodynamical simulations of galaxy evolution with stellar feedback (e.g., \citealp{naggal}, \citealp{Stey23}, \citealp{secgal}) are ultimately the aim, these are computationally costly.
    Multiple dynamical friction tests for the presence of cosmologically relevant dark matter show it not to be present (\citealp{Hernandez25}). For example, the observationally implied orbit of the Small about the Large Magellanic Cloud (\citealp{Massana22}) is not possible if both galaxies have dark matter halos (\citealp{Oehm24};  \citealp{Hernandez25}). Here we therefore assume dark matter halos not to be present which implies that galaxies rarely suffer mergers, and we treat galaxies as evolving largely in isolation as self-regulated systems (e.g., \citealp{Disney08}).
    It allows a rapid assessment based on a parameter scan of how the stellar population, as described by a metallicity- and density-dependent galaxy-wide IMF (gwIMF), self-regulates the evolution of a galaxy.\par
    A proven and useful description is the 2-phase and 2-component model, in which stars, stellar remnants, hot gas and cold gas form a galactic network that can approximately simulate the chemical evolution of a galaxy (\citealp{selfreg}, \citealp{coneva}). Additionally equipped with a (constant) accretion rate such a system shows a self-regulated star-formation process (\citealp{accreg}). For example, from a detailed analysis of field stars based on Gaia data the star-formation history (SFH) of the Milky Way is seen to be flat at about 2 M$_\odot/$yr with modulations, most likely owing to the repeated orbits of the Large Magellanic Cloud and of the Sagittarius satellite galaxy (Fig. 11 in \citealp{Gellert24}; see also Fig. 5 in \citealp{Fernandez25}).\par
    Most of these models are based on an invariant initial mass function (IMF), which is a crucial part of the star-formation description. However, recent studies indicate a variable gwIMF, sensitive to both the SFR and the metallicity of the galaxy (e.g., \citealp{lee}, \citealp{guna}, \citealp{Jerabkova25}, \citealp{KroupaGjergo26}).\par
    Therefore, it is under debate how to formulate an accurate gwIMF for galaxies in general. The approach implemented here in the chemical evolution model is the galaxy-wide stellar initial mass function (IGIMF) theory, a semi-empirical approach to calculate the gwIMF, starting with the stellar IMF on the embedded star cluster scale and building up from this a variable IGIMF. This theory was first introduced by \cite{igimf}, in its modern form by \cite{metsfr}, with the most advanced formulation being available in \cite{Yan21}, \cite{Hasl24}, \cite{Zonoozi25} and \cite{Gjergo26}.
    To better understand the evolution of galaxies, it hence seems useful to compare the results of the 2-phase 2-component model with the canonical (invariant) gwIMF (\citealp{Kroupa2001}) and the Salpeter (invariant) gwIMF (\citealp{Sally}) with the results of the same model, implemented with a variable gwIMF derived from the IGIMF theory.\par
    Calculations with a Salpeter gwIMF are included since it is used in the previous models (see \citealp{selfreg}), otherwise it is only of academic interest.\par
    The main varying input parameters of the models are the effective radius, $R$, the final age of the galaxy, $\tau_{\rm gal}$ and the accretion rate, $A_{g_\mathrm{c}}$.
    All other input parameters are listed in Table~\ref{tab:input}.\par
    Some simplifications of the model are addressed in Sect.~\ref{sec:limits}.

\section{Model}
\subsection{IGIMF}\label{sec:igimf}

The basic principle of the IGIMF theory is simple, but physically motivated. Simply speaking, it states that the IGIMF is the sum of all IMFs in all embedded clusters forming in the molecular clumps over their lifetime, $\delta t \approx 10$ Myr, of their molecular cloud. Hence, the systematic variations in the IGIMF are accounted for through the variation of the stellar IMF on a molecular clump scale, or in other words on the embedded star cluster scale. Other than that, the theory is based on a few well-motivated assumptions, which are the following (for comparison see also \citealp{axioms}, \citealp{galimf}, \citealp{metsfr}, \citealp{chemevo}, \citealp{Yan24}, \citealp{Hasl24}, \citealp{Zonoozi25}):\par
1. Every star forms in a molecular clump, and therefore belongs to an embedded star cluster (which can dissolve over time, \citealp{Kroupa95a}, \citealp{Kroupa95b}, \citealp{lada}, \citealp{megeath}).\par
2. The initial mass distribution of stellar masses in a star cluster follows a stellar IMF. The shape of this stellar IMF is dependent on the 
clump density at birth as well as on the metallicity (\citealp{marks}). There is a relation between the mass of stars in the embedded cluster and the mass of the most massive star in it (\citealp{Yan23}).\par
3. The mass distribution of embedded star clusters follows the embedded star-cluster mass function (ECMF, \citealp{igimf}). The ECMF is an observationally estimated single power law function with a SFR-dependent power-law index $\beta$ between 1.5 and 2.5.\par
4. As already mentioned, at any time $t$ over the time interval $\delta t$ the IGIMF is the sum of all stars in all embedded clusters and can therefore be written as:
\begin{equation}
    \xi_{{\rm IGIMF}}({m,S\!F\!R,t}) = \int_{M_{\rm ecl,min}}^{M_{\rm ecl,max} (\it S\!F\!R)}\xi_{\ast}({m,M_{\rm ecl}})\;\xi_{{\rm ecl}}( {M_{\rm ecl},S\!F\!R,t}) \; dM,
	\label{eq:IGIMF}
\end{equation}
where the stellar IMF is defined as $\xi_{\ast}({m})=\frac{dN_{\ast}}{dm} \propto m^{-\alpha} $ and the ECMF is $\xi_{{\rm ecl}}({M}_{\rm ecl})=\frac{dN_{\rm ecl}}{dM_{\rm ecl}} \propto M^{-\beta (S\!F\!R)}$. Here $dN_{\ast}$ is the number of stars in the mass interval $m$ to $m+dm$ and $dN_{\rm ecl}$ is the number of embedded clusters in the stellar mass interval $M_{\rm ecl}$ to $M_{\rm ecl} + dM_{ecl}$. The power-law index $\alpha$ of the stellar IMF is a function of the stellar mass, the gas-metallicity and density (\citealp{marks}) with the Salpeter value being $\alpha = 2.35$ and with values of $\alpha_1 = 1.3$ for $ 0.1 < m/ \it M_{\odot} < 0.5$ and $\alpha_2 = 2.3$ for $ 0.5 \le m/ \it M_{\odot} \le m_{\rm max}$ for the invariant canonical stellar IMF. In the models used here a stellar mass range from 0.08 $M_{\odot}$ to 150 $M_{\odot}$ is assumed.

$M_{\rm ecl, min} = 5$ $M_{\odot}$ is the mass of the least massive embedded cluster corresponding to about 5 stellar binaries and $M_{\rm ecl, max}(S\!F\!R)$ is the mass of the most massive cluster given the SFR of the galaxy. \par
5. The shortest time period in which enough embedded clusters can form to completely populate the ECMF is $\delta t = 10$\,Myr and in this time the SFR is assumed to be constant (\citealp{weidner}).
From this the total mass in stars of all embedded clusters formed during this period in the galaxy follows as:

\begin{equation}
    M_{\rm tot, \delta  \it t} = S\!F\!R \cdot \delta t
    \label{eq:Mtotgalimf}
\end{equation}

\noindent To implement the IGIMF theory into a chemical evolution model the publicly available python code \texttt{GalIMF}\footnote{The code is available on \href{https://github.com/Azeret/galIMF}{https://github.com/Azeret/galIMF}, here version galimf 1.1.10 is used} is used. In \cite{galimf} also a more detailed description of the IGIMF theory is available. We refer the reader to \cite{Hasl24} for the photGalIMF version which allows the photometric properties of an evolving galaxy to be computed, while the SPS-VarIMF code by \cite{Zonoozi25} allows spectral synthesis thereof.

\subsection{The 2-phase and 2-component model}
\label{sec:phase_comp}

The basis of the model used here was developed by \cite{selfreg} by analyzing a closed system. The extension to a 2-phase 2-component model including accretion was developed in \cite{accreg} and is used here.
The two components are massive stars and the intermediate/low-mass stars plus remnants. The two phases are cold/warm and hot gas. The energetic and matter interactions between the components and phases are depicted in \cite{accreg}, their Figure 1.\par
The simplified model is described by the time derivatives of the components and phases, given by Eqs.~(\ref{eq:cdot}) -- (\ref{eq:rdot}) (see also \citealp{selfreg}):

\begin{equation}
\dot{g_\mathrm{c}} = -\psi -E_{g_\mathrm{c}}+K_{g_\mathrm{h}}+ A_{g_\mathrm{c}},
	\label{eq:cdot}
\end{equation}

\begin{equation}
\dot{g_\mathrm{h}} = \frac{s \cdot \eta}{\langle \tau_{s} \rangle} +E_{g_\mathrm{c}}-K_{g_\mathrm{h}} - \frac{s \cdot \eta_{\rm esc}}{\langle \tau_{s} \rangle},
	\label{eq:gdot}
\end{equation}

\begin{equation}
\dot{s} = \zeta \cdot \psi - \frac{s}{\langle \tau_{s} \rangle},
	\label{eq:sdot}
\end{equation}

\begin{equation}
\dot{r} = (1 - \zeta) \cdot \psi - \frac{(1 - \eta) \cdot s}{\langle \tau_{s} \rangle} -\frac{s \cdot \eta_{\rm esc}}{\langle \tau_{s} \rangle}.
	\label{eq:rdot}
\end{equation}

\noindent Here $g_\mathrm{c}$, $g_\mathrm{h}$, $s$ and $r$ are, respectively, the volume mass densities of the cold and hot gas phase, and of the high-mass stars and remnants/low-mass stars. $\psi$ is the star-formation rate density (SFRD), $E_{g_\mathrm{c}}$ is the evaporation rate, $K_{g_\mathrm{h}}$ the condensation rate and $A_{g_\mathrm{c}}$ the accretion rate, which is held constant. We assume $A_{g_\mathrm{c}}$ to be constant based on the observational result that the SFRs are constant (\citealp{Kroupa20}). This is explicitly evident in the star-formation histories of the Magellanic Clouds (\citealp{Massana22}) and isolated dwarf galaxies (\citealp{Yang24}), furthermore outflows ($\eta_{\rm esc}$) are neglected, see Sect.~\ref{sec:Dis}. $\zeta$ is the formation fraction of high-mass stars, $\eta$ their gas return fraction and $\langle \tau_{s} \rangle$ their mean lifetime, computed with Eq.~(\ref{eq:tau}). All stars with a mass above 10 $M_{\odot}$ are considered as massive stars. $\eta_{\rm esc}$ is the fraction of returned material by SN II which is energetic enough to be considered as outflows (see Eq.~(\ref{eq:eta_esc}) and Sect.~\ref{sec:outflows}). The term with $\eta_{\rm esc}$ in Eqs.~(\ref{eq:gdot}), (\ref{eq:rdot}), (\ref{eq:zg}) and (\ref{eq:zr}) is set to zero within the main text and its influence is discussed in Sect.~\ref{sec:outflows}.\par
The functional form of the SFRD $\psi$ is taken from \cite{selfreg} (there it is called stellar birth function) and has the functional form:

\begin{equation}
\psi(g_\mathrm{c},T_{g_\mathrm{c}}) = C_n g_\mathrm{c}^n exp(-T_{g_\mathrm{c}}/1000).
	\label{eq:SFR}
\end{equation}
The parameters $C_n$ and $n$ can be arbitrarily chosen and are set to $n$ = 2 and $C_n$=0.55, as demonstrated in \cite{selfreg} (originally also found by \citealp{larson}).

The energy exchange in the system (between the two gas phases) is described by:

\begin{equation}
\begin{split}
\frac{\dot{e}_{g_\mathrm{c}}}{\it M_{\odot}\, \rm Myr^{-3}\, pc^{-1}} = & \, \frac{h_{\gamma}}{\rm Myr^{-3}\,pc^{2}} \cdot \frac{s}{\it M_{\odot}\, \rm pc^{-3}} \\ 
 & \, -\frac{g_\mathrm{c}^{2}}{M^2_{\odot}\, \rm pc^{-6}}\cdot \frac{\Lambda(T_{g_\mathrm{c}},Z)}{\it M_{\odot}^{-1}\, \rm Myr^{-3}\,pc^{5}} \\
& \,- \frac{E_{g_\mathrm{c}}}{\it M_{\odot}\, \rm Myr^{-1}\,pc^{-3}}\cdot \frac{b}{\rm K^{-1}\,Myr^{-2}\,pc^2}\cdot \frac{T_{g_\mathrm{c}}}{\rm K} \\
& \,+ \frac{K_{g_\mathrm{h}}}{\it M_{\odot}\, \rm Myr^{-1}\,pc^{-3}}\cdot \frac{b}{\rm K^{-1}\,Myr^{-2}\,pc^2}\cdot \frac{\Tilde{T}_{g_\mathrm{h}}}{\rm K} \\
& \, - \frac{\psi}{\it M_{\odot}\, \rm Myr^{-1}\,pc^{-3}}\cdot \frac{b}{\rm K^{-1}\,Myr^{-2}\,pc^2}\cdot \frac{T_{g_\mathrm{c}}}{\rm K}\\
& \,+ \frac{b}{\rm K^{-1}\,Myr^{-2}\,pc^2}\cdot \frac{T_{A_{g_\mathrm{c}}}}{\rm K}\cdot \frac{A_{g_\mathrm{c}}}{\it M_{\odot}\, \rm Myr^{-1}\,pc^{-3}}\\
& \,+ \frac{1}{2} \frac{v_{\rm ac}^2}{\rm Myr^{-2} pc^2}\cdot \frac{A_{g_\mathrm{c}}}{\it M_{\odot}\, \rm Myr^{-1}\,pc^{-3}},
\end{split}
\label{eq:ecdot}
\end{equation}

\begin{equation}
\begin{split}
\frac{\dot{e}_{g_\mathrm{h}}}{\it M_{\odot}\, \rm Myr^{-3}\, pc^{-1}} = & \, \frac{h_{\rm SN}}{\rm Myr^{-3}\,pc^{2}} \cdot \frac{s}{\it M_{\odot}\, \rm pc^{-3}} \\ 
 & \, -\frac{g_\mathrm{h}^{2}}{M^2_{\odot}\,\rm pc^{-6}}\cdot \frac{\Lambda(T_{g_\mathrm{h}},Z)}{\it M_{\odot}^{-1}\,\rm Myr^{-3}\,pc^{5}} \\
& \,- \frac{E_{g_\mathrm{c}}}{\it M_{\odot}\, \rm Myr^{-1}\,pc^{-3}}\cdot \frac{b}{\rm K^{-1}\,Myr^{-2}\,pc^2}\cdot \frac{\Tilde{T}_{g_\mathrm{c}}}{\rm K} \\
& \,+ \frac{K_{g_\mathrm{h}}}{\it M_{\odot}\, \rm Myr^{-1}\,pc^{-3}}\cdot \frac{b}{\rm K^{-1}\,Myr^{-2}\,pc^2}\cdot \frac{T_{g_\mathrm{h}}}{\rm K} \\
& \, + \frac{b}{\rm K^{-1}\,Myr^{-2}\,pc^2}\cdot \frac{T_{g_\mathrm{h}}}{\rm K} \cdot \frac{s}{\it M_{\odot}\, \rm pc^{-3}} \cdot \frac{\rm Myr}{\langle \tau_{s} \rangle},
\end{split}
\label{eq:egdot}
\end{equation}

\noindent where $h_{\gamma}$ is the heating coefficient for the cool gas, due to ionizing radiation, $h_{\rm SN}$ the respective coefficient for the hot gas caused by supernovae type II, $T_{g_\mathrm{c}}$ the temperature of the cool gas clouds, $T_{g_\mathrm{h}}$ that of the hot gas, and $T_{A_{g_\mathrm{c}}}$ the temperature of the infalling gas, while $v_{\rm ac}$ is the infall velocity so that the last term in Eq.~(\ref{eq:ecdot}) represents the completely dissipated kinetic energy of the infalling gas. $\Tilde{T}_{g_\mathrm{h}}$ and $\Tilde{T}_{g_\mathrm{c}}$ are the temperatures of the corresponding gas entering the cold gas (via condensation) and, respectively, the hot gas (via evaporation). $\Lambda(T,Z)$ is the cooling function that is described in Sect.~\ref{sec:cool}. Finally, $b = 1.5 k_{\rm b}/\mu_{\rm p}$. $\mu_{\rm p}$ is the proton mass and $k_{\rm b}$ the Boltzmann constant.\par
Following \citet{accreg}, in the case of a constant accretion rate, the SFRD will regulate itself and settle in an equilibrium. This equilibrium SFRD can be expressed as:

\begin{equation}
\psi = \frac{1}{1-\eta\zeta} A_{g_\mathrm{c}},
    \label{eq:eqSFR}
\end{equation}

\noindent by setting $\dot{g_\mathrm{c}}=\dot{s}=\dot{g_\mathrm{h}}=0$. As one can see in Eq.~(\ref{eq:eqSFR}), the self-regulated SFRD for an invariant gwIMF is entirely determined by the accretion rate $A_{g_\mathrm{c}}$ and completely independent of the chosen form of the SFRD function. In the case of the IGIMF, $\eta$ and $\zeta$ are also varying, therefore the equilibrium SFRD should be different from the case with an invariant gwIMF, and also depend on the metallicity and thus shift with time.

\subsection{IMF-dependent parameters}
\label{sec:imf-param}

For the implementation of the IGIMF theory the python module \texttt{GalIMF} (\citealp{galimf}) was used. In order to compare the IGIMF model with the ones using an invariant gwIMF a galaxy size had to be assumed in order to convert the SFRD from the evolution model to a SFR, since it is required as an input for the \texttt{GalIMF} code.\par
All IMF-dependent parameters of the model are computed with the invariant gwIMFs as well as with an IGIMF:

\begin{equation}
\zeta = \frac{\int_{10 M_{\odot}}^{m_{\rm max}} m \cdot \xi_{{\rm IGIMF}} \;dm}{M_{\rm tot}},
    \label{eq:zeta}
\end{equation}

\begin{equation}
h_{\gamma} = \eta_{\rm LyC} E_{\rm LyC} \frac{\int_{10 M_{\odot}}^{m_{\rm max}} L_{\rm LyC} \cdot \xi_{{\rm IGIMF}} \;dm}{\int_{10 M_{\odot}}^{m_{\rm max}} m \cdot \xi_{{\rm IGIMF}} \;dm},
    \label{eq:h_ph}
\end{equation}

\begin{equation}
h_{\rm SN} = E_{51} \cdot \int_{m(t+\Delta t)}^{m(t)} \xi_{{\rm IGIMF}} \;dm,
    \label{eq:h_SN}
\end{equation}

\begin{equation}
\eta = \frac{\zeta - m_{{\rm rem}} \int_{10 M_{\odot}}^{m_{\rm max}} \xi_{{\rm IGIMF}} \;dm}{\zeta},
    \label{eq:eta}
\end{equation}

\begin{equation}
\eta_{\rm esc} = \frac{\zeta - m_{{\rm rem}} \int_{10 M_{\odot}}^{M_{\rm esc}} \xi_{{\rm IGIMF}} \;dm}{\zeta},
    \label{eq:eta_esc}
\end{equation}

\begin{equation}
\langle \tau_{s} \rangle = \frac{\int_{10 M_{\odot}}^{m_{\rm max}} \tau_{\ast}(m) \cdot \xi_{{\rm IGIMF}} \;dm}{\int_{10 M_{\odot}}^{m_{\rm max}} \xi_{{\rm IGIMF}} \;dm}.
    \label{eq:tau}    
\end{equation}

\noindent Here, $h_{\gamma}$ is the heating coefficient due to ionizing radiation and is computed via Eq.~(\ref{eq:h_ph}), where $E_{\rm LyC} \approx 13.6$\,eV is the average photon energy, $L_{\rm LyC}$ is the stellar Lyman continuum luminosity (\citealp{selfreg}) and $\eta_{\rm LyC}$ the efficiency factor for the conversion of absorbed photon energy into thermal energy. For the model we assumed $\eta_{\rm LyC} = 10^{-4}$, while \cite{Hensler07} derives for the thermal, kinetic and ionization efficiencies of 15-85 $M_{\odot}$ stars values around $10^{-3}$.\par
The heating due to type II supernovae (SN II) is given by Eq.~(\ref{eq:h_SN}) (\citealp{SNII}), where $E_{51}$ is the energy injected into the gas by a supernova, which is around $10^{51}$\,erg and the number of SN II during a given period $\Delta t$ is given by $N_{\rm {SN II}}(t, \Delta t) = \int_{m(t+\Delta t)}^{m(t)} \xi_{{\rm IGIMF}} \;dm , \; m> 10 \rm\; M_{\odot}$. We adopt the heating efficiency of 100\,\% in Eq.~(\ref{eq:h_SN}) following \cite{SNII} but note that 1-10\,\% would be more realistic.\par
Note that a higher value of any of the two heating coefficients leads to a stronger self-regulation, and therefore a longer star-formation period.\par
In the model, the average mass of a remnant $m_{\rm rem}$, is assumed to be 1.4 $M_{\odot}$. To compute the mass-weighted mean lifetime of massive stars, $\langle \tau_{s} \rangle$, the mass-lifetime relation from \cite{stellarlifetime} $\tau_{\ast}(m_{\ast})=50/10^{0.6} \;(m_{\ast}/10 M_{\odot})^{-0.6}\; \text{Myr}$ was used in Eq.~(\ref{eq:tau}).\par

As mentioned in Sect.~\ref{sec:phase_comp}, $\eta_{\rm esc}$ is the fraction of returned gas from SN II explosions that is expelled from the galaxy as outflows. The upper integration limit of Eq.~(\ref{eq:eta_esc}), $M_{\rm esc}$, is computed as:

\begin{equation}
M_{\rm esc} = m_{\rm rem} + \frac{2\cdot E_{51}}{v_{\rm esc}^2},    
    \label{eq:M_esc}
\end{equation}

\noindent with

\begin{equation}\label{eq:vesc}
    v_{\rm esc} = \sqrt{\frac{2GM_{\rm tot}}{R}},
\end{equation}

\noindent where $g_\mathrm{h}$ is the Gravitational constant, $R$ the effective radius and $M_{\rm tot}$ the total mass of the galaxy within $R$.\par
We note that Type Ia SNe are not included in the present formalism because modeling their delayed enrichment requires an extended framework with additional stellar components and a delay-time distribution. As a consequence, the metallicity $Z$ in our model primarily traces $\alpha$-element–like enrichment from core-collapse supernovae.\par
However, the main focus of this work is the self-regulation of star-formation and in this context Type Ia SNe should have a minor influence, since the galaxy-wide regulation is driven mainly by SN II. We also estimate the error on the global metallicity $Z$ by assuming around 0-80\,\% of iron is produced by Type Ia SNe (\citealp{Wiersma09}, \citealp{Palicio23}) and calculating the fraction of the iron yield and the total metallicity yield in our model. From this we estimate the error on our global $Z$ as:

\begin{equation}\label{eq:Z_error}
    \epsilon_Z = \frac{f_{\rm Fe,Ia}}{1-f_{\rm Fe,Ia}} \cdot \frac{y_{ \rm Fe}}{y}.
\end{equation}

\noindent $f_{\rm Fe,Ia}$ is the fraction of iron produced by Type Ia SNe, $y_{ \rm Fe}$ is the SN II iron yield and $y$ the total SN II metal yield for our models (see Sect.~\ref{sec:metal}). While $f_{\rm Fe,Ia}$ is largest for metal-rich, late-time systems, the yield ratio increases in systems with lower high-mass star-formation rates, where the total metal yield is reduced relative to iron. Overall we estimate the error based on Eq.~(\ref{eq:Z_error}) to be between 1 and 20\,\% for the IGIMF model and even lower for the models with a canonical IMF. These numbers are of course approximate. We emphasize that astrophysically correct chemical evolution studies must include the contributions by Type Ia SNe and that here we are merely checking how the variable vs a universally valid invariant gwIMF affects the overall metallicity which is dominated by the contributions from core collapse supernovae. In this context it is reasonable for our approach to only taking SN II into account.

All of the parameters are also computed using a Salpeter IMF with a slope $\alpha$ = 2.35 and lower and upper mass limits of 0.1 $\text{M}_{\odot}$ and 100 $\text{M}_{\odot}$ respectively. For the massive-stars fraction this formation yields a value of $\zeta$ = 0.12 and for the gas return fraction $\eta$ = 0.94 (\citealp{selfreg}). The mass-weighted mean lifetime of massive stars for a Salpeter IMF is $\approx$ 9 Myr.\par

For the two heating coefficients the values $h_{\gamma}$ = 25840 pc$^2$ Myr$^{-3}$, $h_{\rm SN}$ = 290273 pc$^2$ Myr$^{-3}$ were used in cases with a Salpeter IMF (Eq.~(\ref{eq:h_ph}), (\ref{eq:h_SN}), respectively). For the canonical IMF (\citealp{Kroupa2001}), the coefficients are $\zeta$ = 0.19, $\eta$ = 0.946, $h_{\gamma}$ = 85134 pc$^2$ Myr$^{-3}$ and $h_{\rm SN}$ = 389895 pc$^2$ Myr$^{-3}$. Note that in the case of the invariant stellar IMFs star clusters can host fractions of massive stars for low SFRs (see \citealp{Stey23} for a discussion).

\subsection{Metallicity}
\label{sec:metal}

The time evolution of the metal mass volume densities, $z_i$, in each component is computed based on \cite{coneva}. The $z_i$ in this framework are the product of the metallicity $Z_i$ and the mass volume density of corresponding component $i = g_\mathrm{c},g_\mathrm{h},s,r$ (e.g., $z_s = Z_s \cdot s$ and $\dot{z_s} = \frac{d}{dt}(s \cdot Z_s)$). Explicitly the four equations are:

\begin{equation}
\dot{z}_{g_\mathrm{h}} = \frac{s}{\langle \tau_{s} \rangle}\left( \eta Z_s + y\frac{1-\zeta \eta}{\zeta} (1-Z_s)\right) + E_{g_\mathrm{c}}\cdot Z_{g_\mathrm{c}} -K_{g_\mathrm{h}}\cdot Z_{g_\mathrm{h}}  -\frac{s \cdot \eta_{\rm esc}}{\langle \tau_{s} \rangle} \cdot Z_s,
    \label{eq:zg}    
\end{equation}

\begin{equation}
\dot{z}_{g_\mathrm{c}} = -\Psi Z_{g_\mathrm{c}} - E_{g_\mathrm{c}}\cdot Z_{g_\mathrm{c}} +K_{g_\mathrm{h}}\cdot Z_{g_\mathrm{h}} +A_{g_\mathrm{c}} \cdot Z_A,
    \label{eq:zc}
\end{equation}

\begin{equation}
\dot{z}_s = \zeta \Psi Z_{g_\mathrm{c}} - s Z_s/\langle \tau_{s} \rangle,
    \label{eq:zs}
\end{equation}

\begin{equation}
\dot{z}_r = (1-\zeta) \Psi Z_{g_\mathrm{c}} + (1-\eta) s Z_s/\langle \tau_{s} \rangle -\frac{s \cdot \eta_{\rm esc}}{\langle \tau_{s} \rangle} \cdot Z_s.
    \label{eq:zr}
\end{equation}

\noindent In Eq.~(\ref{eq:zg}) for $\dot{z}_{g_\mathrm{h}}$, the first term takes into account the return of present metals to the ISM by massive stars, as well as the return of freshly synthesized metals via the yield $y$. The metals synthesized only by massive stars are given by the yield term, $y/\zeta$ in Eq.~(\ref{eq:zg}).\par
The yield depends on the description of the gwIMF, hence one set of metal yields is used for the calculations with the canonical gwIMF and another for the Salpeter gwIMF, both taken from \citet[their Table 3]{canyield} with an upper mass limit of 100 $M_{\odot}$ and stellar yields from \cite{Nomoto2013}. For the IGIMF model, the yield is computed here using stellar yields given in \cite{Portinari97} and for comparison also with stellar yields from \cite{Nomoto2013} (see Sect.~\ref{sec:yield}). Since the  yield term, $y/\zeta$, in Eq.~(\ref{eq:zg}) is only taking into account the contribution by high-mass stars, the calculated yields for the IGIMF cases based on the stellar yield tables from \cite{Portinari97} and \cite{Nomoto2013} are computed for stars with masses larger than 10 $M_{\odot}$ and used as the term $y/\zeta$. However, the exact description of stellar yields is under debate.\par
Although we refer to the metallicity calculations as chemical evolution, it is worthy to note that the approach here only takes into account the evolution of a single global $Z$-value for each component (see Eqs.~(\ref{eq:zg}) -- (\ref{eq:zr})) and the enrichment is solely driven by high-mass stars. This should be a justified assumption for our calculations, however it is not a full chemical evolution model, for such models see e.g., \cite{canyield}, \cite{Romano17}, \cite{chemevo}, \cite{Gjergo23}.

\begin{figure}
    \includegraphics[width=\columnwidth,height=8cm]{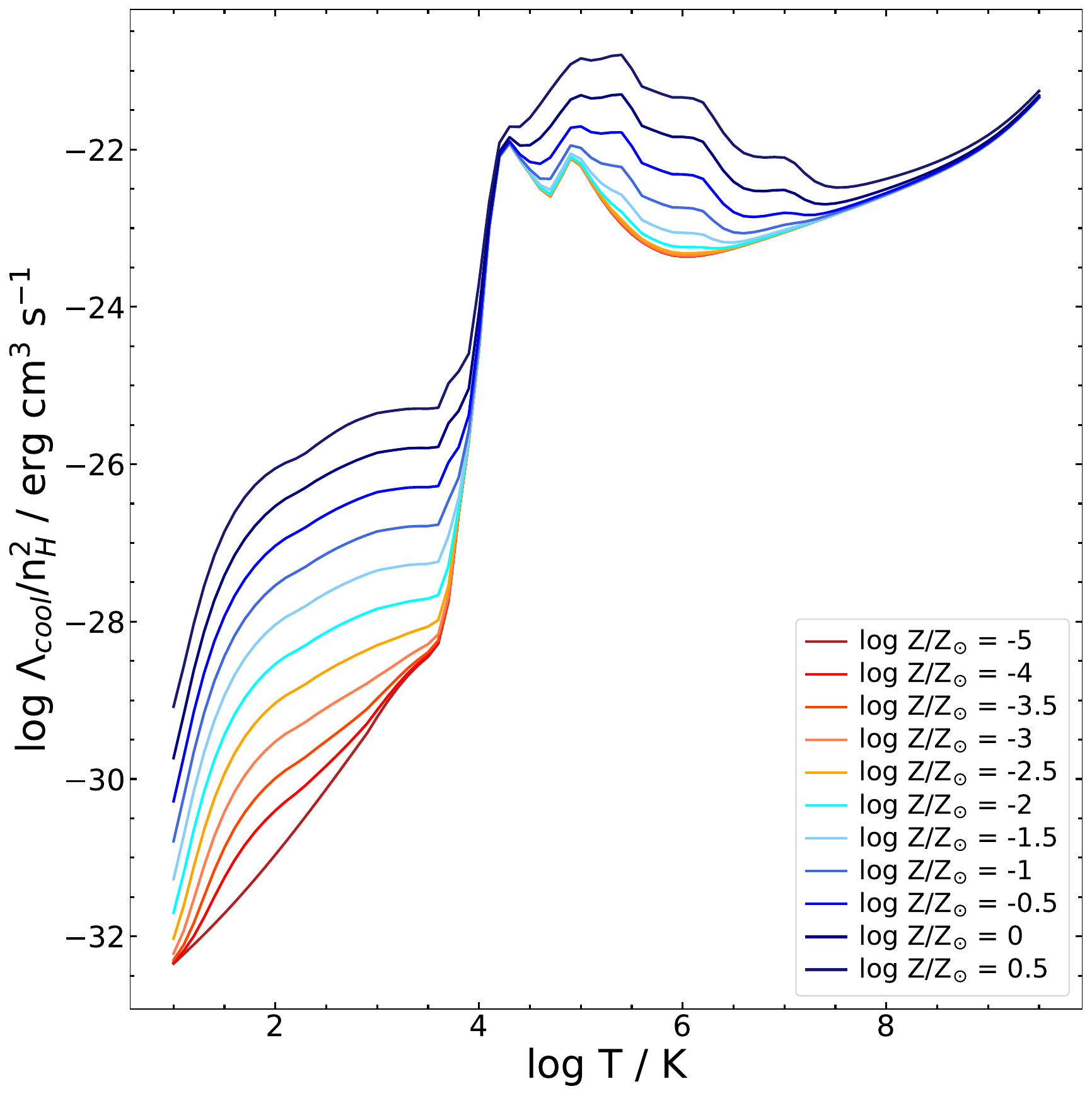}
    \caption{Cooling functions for different metallicities. The redshift was set to zero and the hydrogen particle density to 1 $\text{cm}^{-3}$. The functions were computed using the routine published in \cite{cool} and based on the 2017 release of CLOUDY (\citealp{cloudy}).}
    \label{fig:cool}
\end{figure}

\subsection{Cooling function}
\label{sec:cool}
For each time step the cooling coefficient $\Lambda(T,Z)$ in Eqs.~(\ref{eq:ecdot}) and (\ref{eq:egdot}) is computed for the current temperature and metallicity based on \cite{cool} and their published routine\footnote{\href{https://radcool.strw.leidenuniv.nl/}{https://radcool.strw.leidenuniv.nl/}}. For the computation, the fiducial model of \cite{cool} is used, the redshift is set to zero (although varying the redshift should not influence the cooling function) and the particle density for hydrogen $n_{\rm H}$ was set to 1 $\text{cm}^{-3}$ (this is a low value for cold gas clouds, but since only one cooling function is used for both cold and hot gas phase the choice is reasonable). The resulting cooling curves for different metallicities are shown in Fig.~\ref{fig:cool}.

\subsection{Scope and Limitations}
\label{sec:limits}

It is important to clarify the scope and limitations of the here used approach in order to better interpret the results. The goal of this work is to study gwIMF-driven effects within a reduced physical network. Our galaxy model omits parts of real galaxy networks, for example it has no radial structure or surface-density dependence, no inclusion of AGN or virial-shock heating, no magnetic fields or cosmic rays. Furthermore it omits molecular gas regulation, cloud-cloud collisions and turbulent or gravitational dissipation and stellar winds caused by intermediate stars.\par
It is also important to emphasize that our models evolve with a constant radius, set a priori. Calculations with an evolving radius could improve our models in future studies.\par
Despite these simplifications the model gives a clear and valid insight into gwIMF-driven effects in galactic networks. Furthermore, the simplifications enable a rapid assessment making it possible to study galaxy models over a wide range of parameters.

\section{Calculations}

\label{sec:Sim}
The calculations based on the model described in Sect. \ref{sec:phase_comp} were carried out in Python using a fifth-order Cash-Karp method (\citealp{CashKarp}) with adaptive step size for numerical integration of Eqs.~(\ref{eq:cdot}) -- (\ref{eq:egdot}) and  (\ref{eq:zg}) -- (\ref{eq:zr}). 
For the computation of the IGIMF, the Python code \texttt{GalIMF} (\citealp{galimf}) was used and the intrinsic parameters (see also \citealp{galimf}) of the \texttt{GalIMF} code are listed in Table \ref{tab:IGIMFparam}, all the non-varying initial conditions of the chemical evolution model are listed in Table \ref{tab:inicon}. The input parameters for the code were the SFR within the effective radius $R$, the metallicity of the cold gas phase $Z_c$ and the $\delta t$ (see Eq.~(\ref{eq:Mtotgalimf})) was always 10 Myr for all calculations.\par
The initial density for all models with continuous infall, which should represent star-forming galaxies, is chosen to be the mean baryonic mass density in the universe at redshift z = 10, $\rho_{\rm bar} \approx 8.33 \times10^{-6}$ $M_{\odot}$/pc$^3$.\par
For the calculations that should represent elliptical galaxies, the initial gas densities are calculated from \citet[their Eq.~(9)]{downsizing} for a given final total mass. These calculations also have a short accretion with duration $\tau_{\rm ac}$, as opposed to a constant, lifelong accretion (see Table~\ref{tab:input}). This kind of accretion history is comparable to a monolithic collapse. The accretion time is also taken from \cite{downsizing}.
For all models the gas accretion starts at $t = 0$.\par
The main input parameters specifying the galaxy models in the calculations are then the accretion rate $A_{g_\mathrm{c}}$, the effective radius $R$ (see Sect. \ref{sec:size}), the age of the galaxy $\tau_{\rm gal}$, and the choice of the gwIMF. It is important to note that all quantities are derived within the effective radius. Furthermore, the influence of different stellar yields in the computations with the IGIMF were tested (see Sect.~\ref{sec:yield}), as well as the influence of a different infall velocity (see Sect.~\ref{sec:vac}) or the inclusion of outflows (see Sect.~\ref{sec:outflows}).\par
For each set of initial parameters ($A_{g_\mathrm{c}}$, $R$, $\tau_{\rm gal}$) three models are calculated each with a different assumed gwIMF, one with the invariant Salpeter IMF (model name: SalIMF), one with the invariant canonical IMF (model name: canIMF) and one with the IGIMF (model name: IGIMF). The three initial parameters are fixed over all IMF variants and chosen such that the radius-mass relation and the stellar mass-SFR relations are imposed and all other relations (see Sect.~\ref{sec:Res}) are emergent. All models with their input parameters are listed in Table \ref{tab:input}.\par
The ages, $\tau_{\rm gal}$, of the models are chosen in such a way that the model galaxies fit the stellar-population-mass-SFR relation, with younger ages for lighter model galaxies (see \citealp{galage}). Since the model only contains two stellar components, $s$ and $r$, $M_{\star}$ in the following context always is the stellar-population-mass:
\begin{equation}\label{eq:stellarMass}
    M_{\star} = (s + r) V_{\rm gal}.
\end{equation}
$V_{\rm gal}$ is the volume of the model galaxy that is defined as $V_{\rm gal} =  (4/3)\pi R^3$. Similarly the gas mass $M_{\rm gas}$ is calculated:

 \begin{equation}\label{eq:gasMss}
     M_{\rm gas} = (g_\mathrm{c} + g_\mathrm{h}) V_{\rm gal}.
 \end{equation}

\section{Results}\label{sec:Res}
\begin{figure}
\includegraphics[width=\columnwidth]{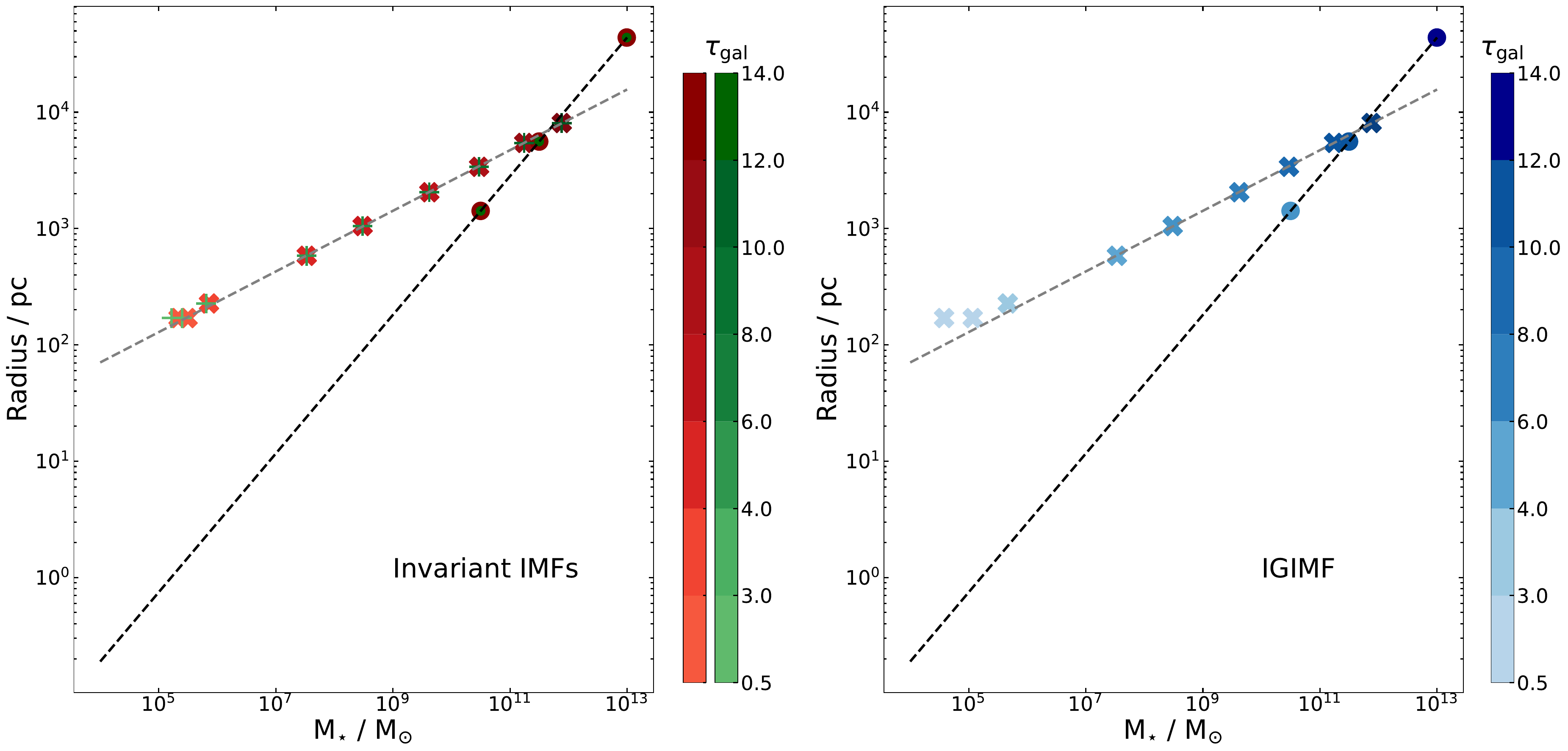}
\caption{Mass-radius relations for all models, left panel shows the ones with a non-varying gwIMF (SalIMF in red and canIMF in green, these overlap), the right panel the calculations with an IGIMF. The x-symbols represent systems with continuous infall and the circles short-duration-infall models. The black dashed line is Eq.~(4) from \cite{size} for pressure supported systems, the gray dashed line is derived from Fig. 2 in \cite{Lelli2016} for star-forming systems, assuming a mass-to-light ratio of $\Upsilon_{\star} = 0.5$ $M_{\odot}$/$L_{\odot}$. The color shade gives the age of the galaxy in Gyr, according to the colorbar.}
\label{fig:size}
\end{figure}

In this section the results and important aspects of the calculations are presented and explained.

\subsection{Radius}\label{sec:size}

The effective radii, $R$, of the calculated systems are chosen such that the resulting final stellar-population-mass and radii fit the stellar mass-radius relation derived from the SPARC galaxy sample (\citealp[their Fig. 2]{Lelli2016}) for star-forming galaxies, using a stellar-population mass-to-light ratio $\Upsilon_{\star} = 0.5$ $M_{\odot}$/$L_{\odot}$. For the galaxies representing elliptical galaxies the mass-radius relation from \citet[their Eq.~(4)]{size} is used.\par
In Fig.~\ref{fig:size} the effective radii are plotted against the final stellar-population-mass for all models, in the left panel the calculations with a non-varying gwIMF are shown and in the right panel the IGIMF model is shown.\par 
The crosses represent calculations with constant infall, which should represent star-forming galaxies\footnote{The star-forming galaxies in the local cosmological volume have had near to constant SFRs over their lifetime (\citealp{Kroupa20}, \citealp{Massana22}, \citealp{Gellert24}, \citealp{Hasl24} and \citealp{Yang24}).\label{footnote:3}}, the dots the calculations with a short and heavy accretion. They should represent quenched or red elliptical galaxies. The age of the model galaxies increases from light colors (youngest) to dark colors (oldest).\par 
The two different non-varying gwIMF models do not differ in their mass and all agree with the relation from \cite{Lelli2016} or \cite{size}.\par
For the IGIMF model only the lightest two galaxies show a slight deviation from the mass-radius relation. Reasons and implications of this deviation at the low-mass end are discussed in Sect. \ref{sec:Dis}. 

\begin{figure}
   \includegraphics[width=\columnwidth]{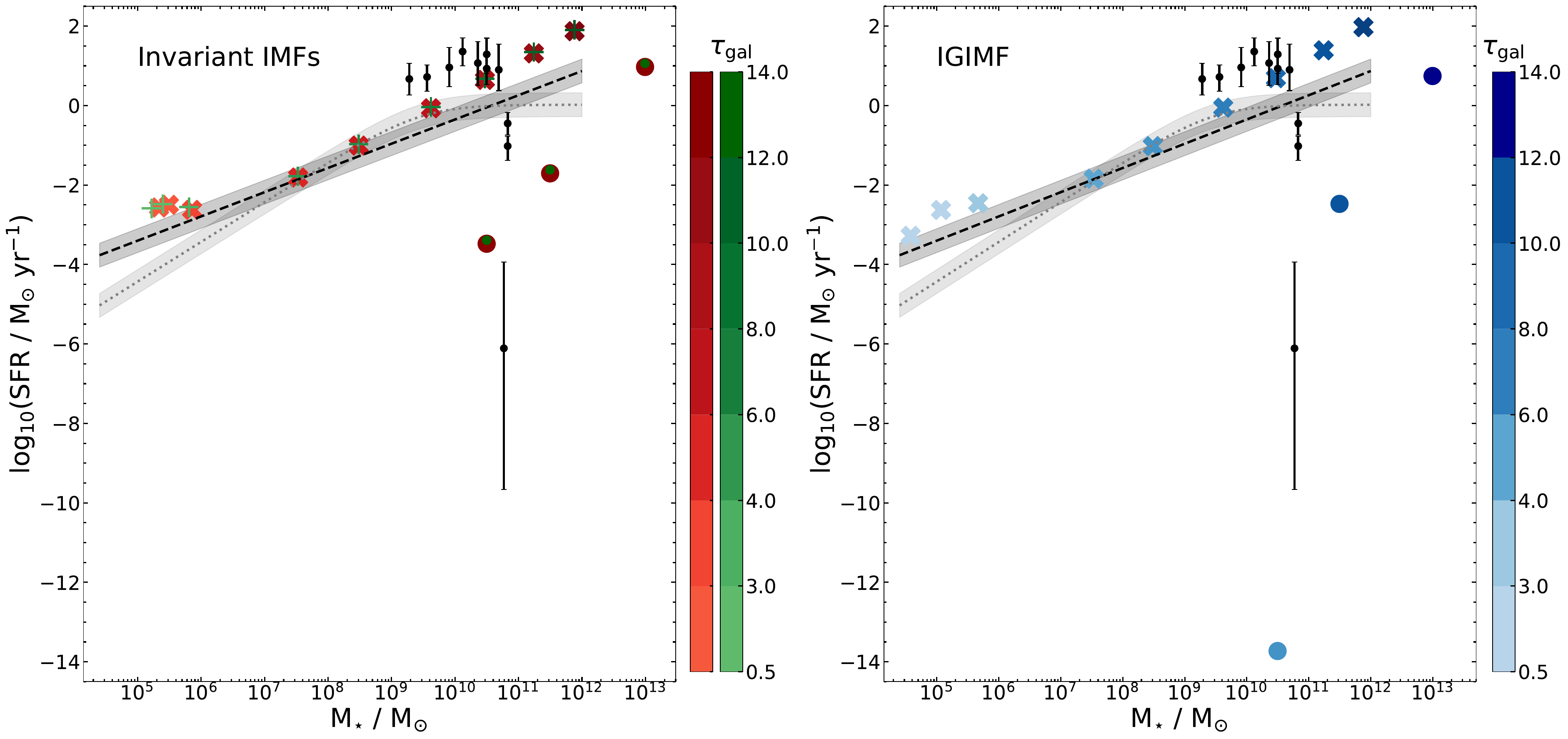}
   \caption{Stellar-population-mass-SFR relation for all models. In the left panel log$_{10}$(SFR) is plotted against the stellar-population-mass for the models SalIMF (red markers) and canIMF (green markers), again the symbols overlap. The right panel shows the same plot for the IGIMF model (blue markers). x-symbols represent continuous accretion models while the circles represent short, heavy accretion models. The black dashed line is the observed stellar-mass-SFR relation from \cite{Hasl24} and the gray dotted line the observed relation from \cite{GMSLeslie20}, the shaded area is the 0.3 dex scatter, which is given as the standard deviation in \cite{GMSLeslie20}. The black data points with error bars depict the observational data from \cite{Siudek18}. The color shade gives the age of the galaxy in Gyr, according to the colorbar.}
    \label{fig:GMS14}%
\end{figure}
\subsection{Stellar mass-SFR relation}\label{sec:GMS}

For all calculations, the final SFR against the stellar-population-mass is plotted in Fig.~\ref{fig:GMS14}. In the left panel, the calculations for a canonical IMF (green) as well as for a Salpeter IMF (red) are plotted, the right panel shows the calculations for an IGIMF. The crosses again represent star-forming galaxies, while for the dots the calculations were done with only a short accretion time $\tau_{\rm ac}$ but higher accretion rate density. These points should represent present-day quiescent elliptical galaxies. The accretion times are calculated from \cite{downsizing} and are 540 Myr, 303 Myr and 128 Myr, with the heaviest galaxy having the shortest accretion time.\par

For all models, the resulting stellar-population-mass-SFR relation or galaxy main sequence (MS) is compared to the observed relations from \cite{Hasl24} and \cite{GMSLeslie20}. For all models the calculated MS lies slightly above that of \cite{Hasl24}, except in the mass regime between $10^8$ and $10^{10}$ $M_{\odot}$, where it is comparable with both relations. However, there is no comparable flattening at higher masses like in \cite{GMSLeslie20}. The circled markers lie significantly below the literature relations for the lightest two models, but agree with the trend depicted by the black data points with error bars, observed by \cite{Siudek18}, which is expected since the galaxy MS is a relation for star-forming galaxies, whereas these calculations should represent quiescent galaxies. The heaviest elliptical model however is not quenching, therefore suggesting the necessity of an additional gas heating source for these types of galaxies.\par 
While no significant difference is visible between the left and right panels for the star-forming galaxies, the lightest elliptical type galaxy calculated with the IGIMF shows a much lower SFR than the comparable calculations with the non-varying gwIMF. This is due to the different feedback and gas-return from a top-heavy respectively top-light gwIMF.\par 
A strong gas infall leads to a starburst. In the IGIMF model this leads to a rapid heating of the cool gas phase, transitioning most of the cool gas to the hot gas phase. Together with the high star-formation rate this leads to a strong decrease of the amount of cool gas. The system then regulates itself and the star-formation rate decreases after a short time. At the same time, the high-mass stars die rather quickly, leading to decreased heating via SN II explosions. However, the SFR decreases to such an extent that the IGIMF becomes top-light, meaning no new massive stars are formed. The system settles into equilibrium, however with a lower amount of cold gas and a higher amount of hot gas than in the non-variable gwIMF cases. The result is more hot gas in the IGIMF case due to the initial high heating from a top-heavy IMF. While in all strong-gas-infall models this hot gas cools and is consumed through star-formation, if the gwIMF is bottom-light, no new gas enters from evolving massive stars (lowest circular point in the right panel of Fig.~\ref{fig:GMS14}). These mechanisms lead to the observed difference of the 'elliptical type' calculation in Fig.~\ref{fig:GMS14}. The star-formation histories are discussed in more detail in Sect.~\ref{sec:self_regualtion}.

\begin{figure}
   \includegraphics[width=\columnwidth]{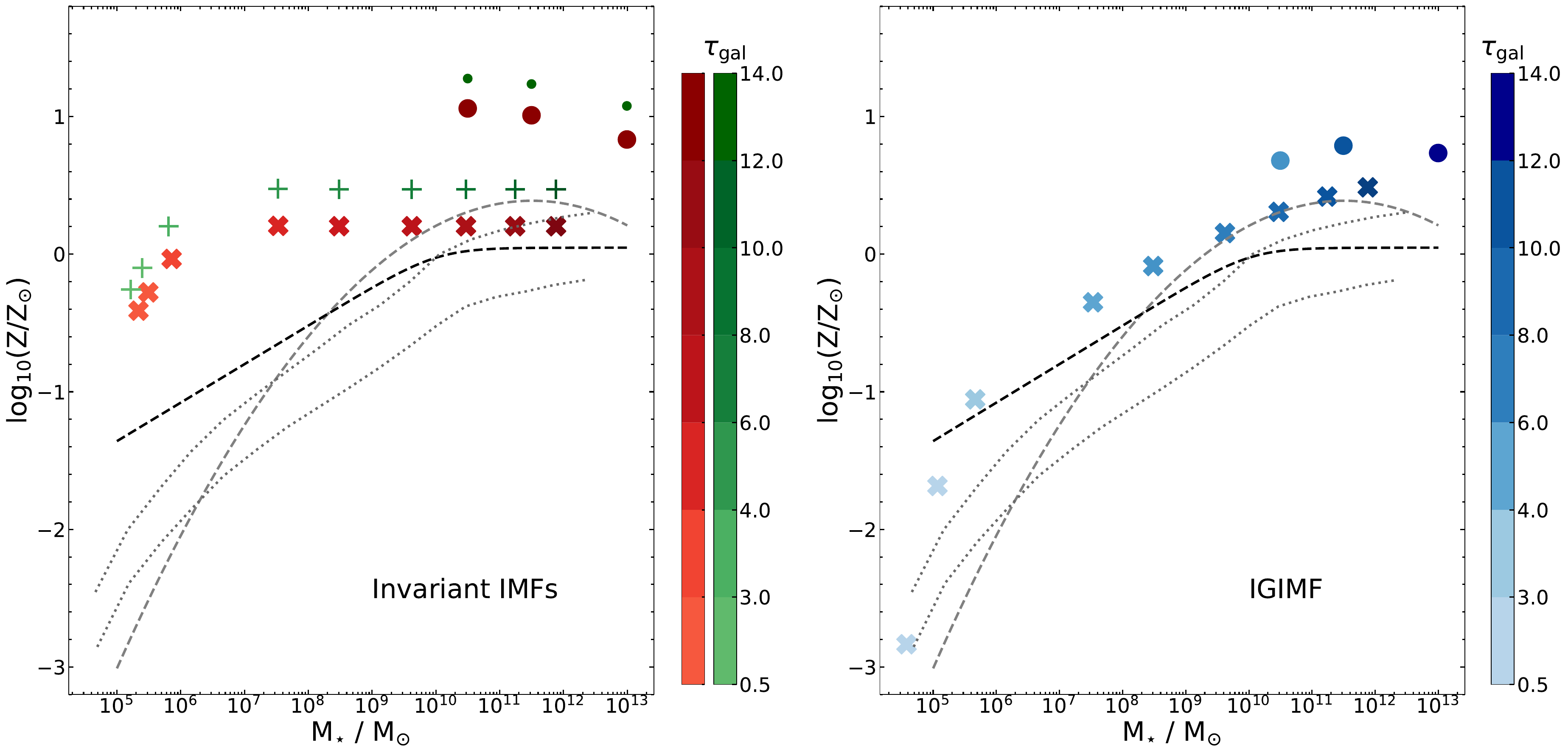}
    \caption{Stellar-population-mass-metallicity relation for all models. In the left panel log$_{10}$(Z/Z$_{\odot}$) is plotted against the stellar-population-mass for the SalIMF (red markers) and the canIMF (green markers) models. The right panel shows the same plot for the IGIMF model (blue markers). x-symbols represent continuous infall models while the circles represent short-heavy accretion models. The black dashed line is the MZR from \cite{MZCurti20}, the gray dashed line the relation from \cite{MZTremonti04} and the dotted lines delineate the relation from \cite{Hasl24}. One can see that the IGIMF models naturally induce a MZR, while for the invariant gwIMFs the metallicity is largely independent of the SFR. The color shade gives the age of the galaxy in Gyr, according to the colorbar.}
    \label{fig:ZM14}%
\end{figure}

\subsection{Mass-metallicity relation}\label{sec:ZM}

Another fundamental relation for galaxies is the relation between stellar mass and gas-phase metallicity. If the calculations should reproduce a realistic galaxy evolution, they also have to reproduce the mass-metallicity relation (MZR). In Fig.~\ref{fig:ZM14}, similar to Fig.~\ref{fig:GMS14}, the gas-phase metallicity is plotted over the stellar-population-mass -- for non-varying gwIMFs in the left panel, and for the IGIMF model in the right panel.\par

For both canIMF and SalIMF, the calculations do not reproduce the MZR given in \cite{MZCurti20} (black dashed line), \cite{MZTremonti04} (gray dashed line), or \cite{Hasl24} (gray dotted line). Rather, the metallicity seems to be independent of the galaxy mass, being slightly above solar metallicity, whereas the canonical IMF causes a higher gas-phase metallicity than the Salpeter IMF because it has a higher fraction of high-mass stars leading to a higher enrichment of the gas. Only the lightest galaxies show a slight decrease in metallicity.\par
A different behavior is visible in the right panel for the IGIMF. The top-lightness of the IMF in low SFR environments (\citealp{metsfr}) leads to a deficit of massive stars, which are the key contributors to heavy elements and hence to an enrichment of the gas. For high-mass galaxies, the calculations reproduce the literature MZR (\citealp{MZTremonti04}) very well. However, for low-mass galaxies the calculated MZR exhibits higher metallicities than expected from \cite{MZTremonti04} and fits better with the other relations. This could also be an effect of the missing inclusion of galactic winds/outflows in the models (see Sect.~\ref{sec:outflows}).\par
It should always be kept in mind that in our model only high-mass stars are responsible for the metal enrichment and that the absolute metallicity depends sensitively on the yield of these massive stars (see Sect.~\ref{sec:metal}). From Fig.~\ref{fig:ZM14} it becomes also clear that different observational MZR calibrations are systematically offset.

\begin{figure}
   \includegraphics[width=\columnwidth]{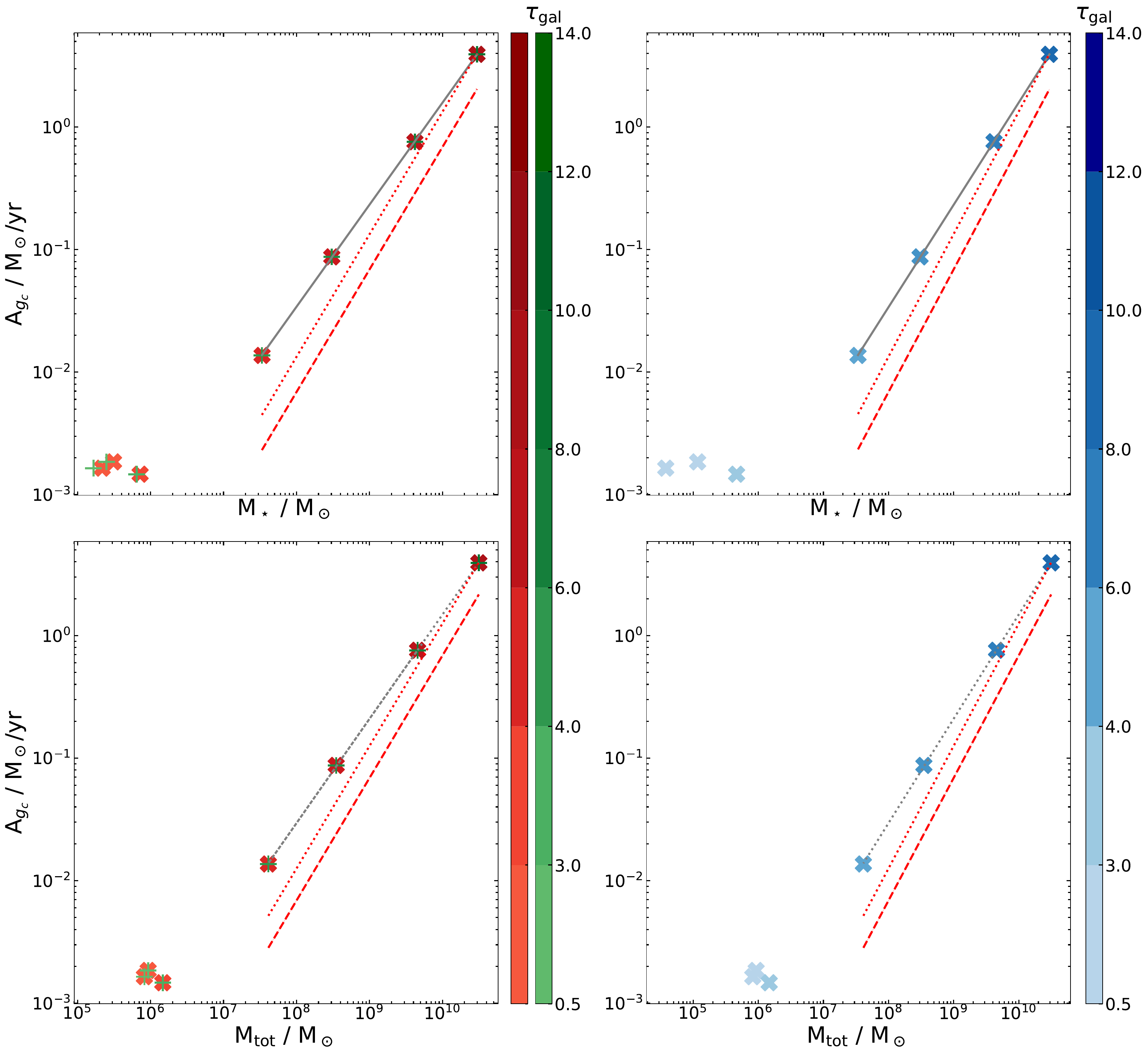}
    \caption{Accretion rate, $A_{g_\mathrm{c}}$, as a function of the stellar-population-mass of the galaxy in the upper panels and as a function of total baryonic mass in the lower panels. The left panels are the same calculations as in Fig.~\ref{fig:GMS14} and in Fig.~\ref{fig:ZM14}. The models SalIMF and canIMF have almost identical results, therefore the points overlap. In the right panels the according IGIMF calculations (see again Fig.~\ref{fig:GMS14} or Fig.~\ref{fig:ZM14}) are shown. The solid gray lines depict the result from a logarithmic least-squares regression fit, the Pearson correlation coefficient is 0.9957 for the canIMF model fit and 0.9934 for the IGIMF model (see Eqs.~(\ref{eq:canFIT}) and (\ref{eq:IGIMFFIT})) if the fit is done for the stellar-population-mass of the galaxy, and also the lightest three galaxies are ignored. The gray dotted lines show the relation between $A_{g_\mathrm{c}}$ and the total mass, M$_{\rm tot}$ = M$_{\rm gas}$ + M$_{\star}$ instead of the stellar-population-mass, which changes the Pearson correlation coefficient to 0.9985 in both cases. In all four panels the red dashed lines correspond to a linear relation between $A_{g_\mathrm{c}}$ and $M$ (see Eq.~(\ref{eq:H0_fit})) and a slope equal to the Hubble constant $H_0 = (67.4 \pm 0.5)$ km s$^{-1}$ Mpc$^{-1}$ (\citealp{Planck}). In Eq.~(\ref{eq:H0_fit}), the $H_0$ parameter fixes the red dashed line to the position slightly to the right of the calculated points, a different $H_0$ value would change the position. The red dotted line is a linear fit to the M$_{\star}$ data points (upper panels) and to the M$_{\rm tot}$ data points (lower panel) (with Eq.~(\ref{eq:H0_fit})) with a fitted $H_0$ parameter. This fitted $H_0$ values for both the canIMF and the IGIMF model is $H_0 = (123.045373 \pm 0.00001)$ km s$^{-1}$ Mpc$^{-1}$ if fitted to the total baryonic mass. If only the stellar-population-mass is considered the fitted values are $H_0 = (130.57437 \pm 0.00001)$ km s$^{-1}$ Mpc$^{-1}$ for the canIMF model and $H_0 = (130.95571 \pm 0.00001)$ km s$^{-1}$ Mpc$^{-1}$ for the IGIMF model. The color shade gives the age of the galaxy in Gyr, according to the colorbar.}
    \label{fig:M_Ac}%
\end{figure}

\subsection{Self-regulation}\label{sec:self_regualtion}
\begin{figure}
   \includegraphics[width=\columnwidth]{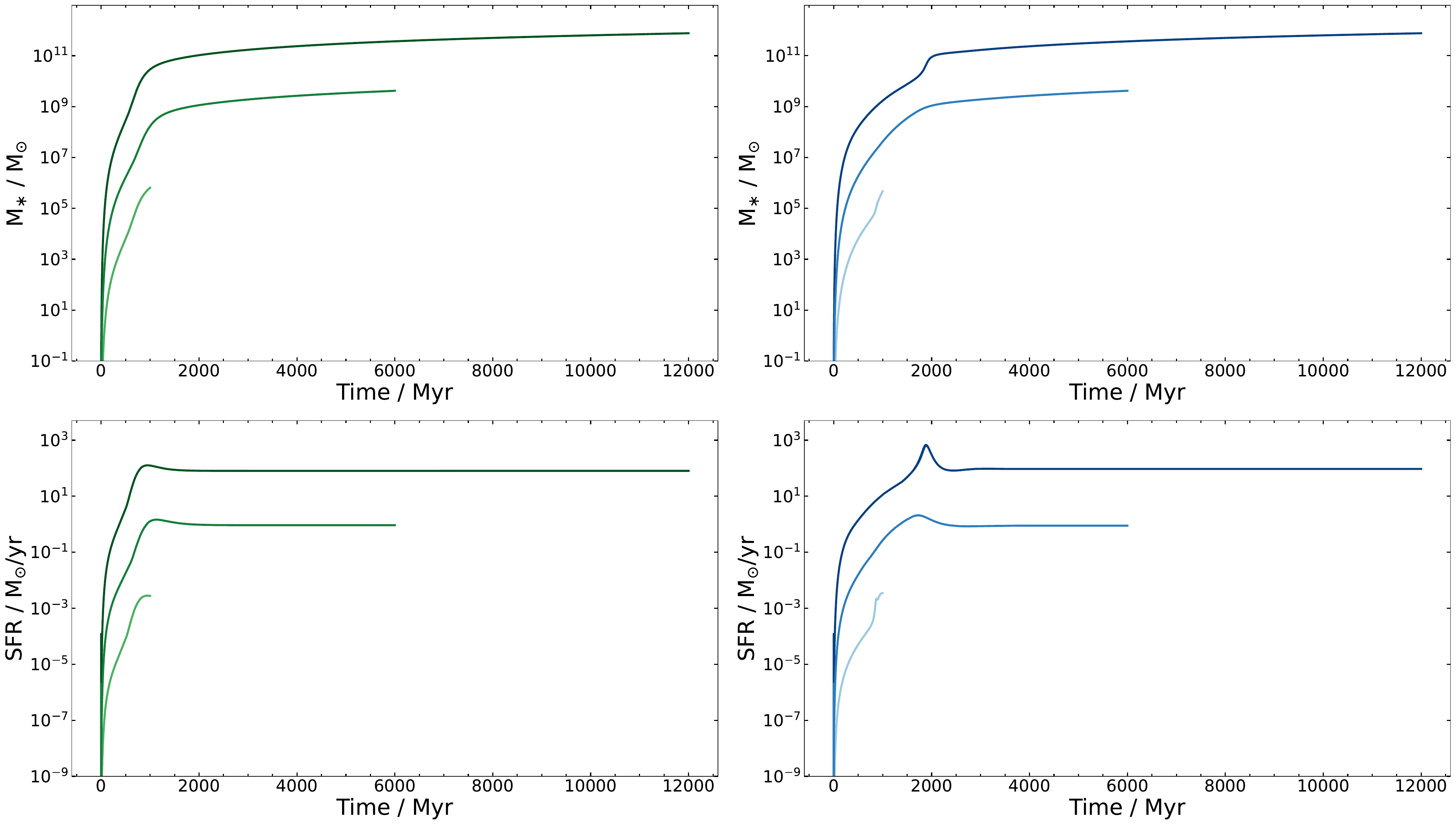}
    \caption{Evolution of the SFR (bottom panel) and stellar-population-mass (top panel) of three different sized model galaxies for the canIMF (green) and the IGIMF (blue) model, all with a constant accretion. The final masses of the model galaxies are $10^{5.8}$, $10^{9.6}$ and $10^{11.9}$ $M_{\odot}$ for the canIMF models and $10^{5.7}$, $10^{9.6}$ and $10^{11.9}$ $M_{\odot}$ for the IGIMF models. The darker the color the more massive is the galaxy.}
    \label{fig:evol}%
\end{figure}

\begin{figure}
   \includegraphics[width=\columnwidth]{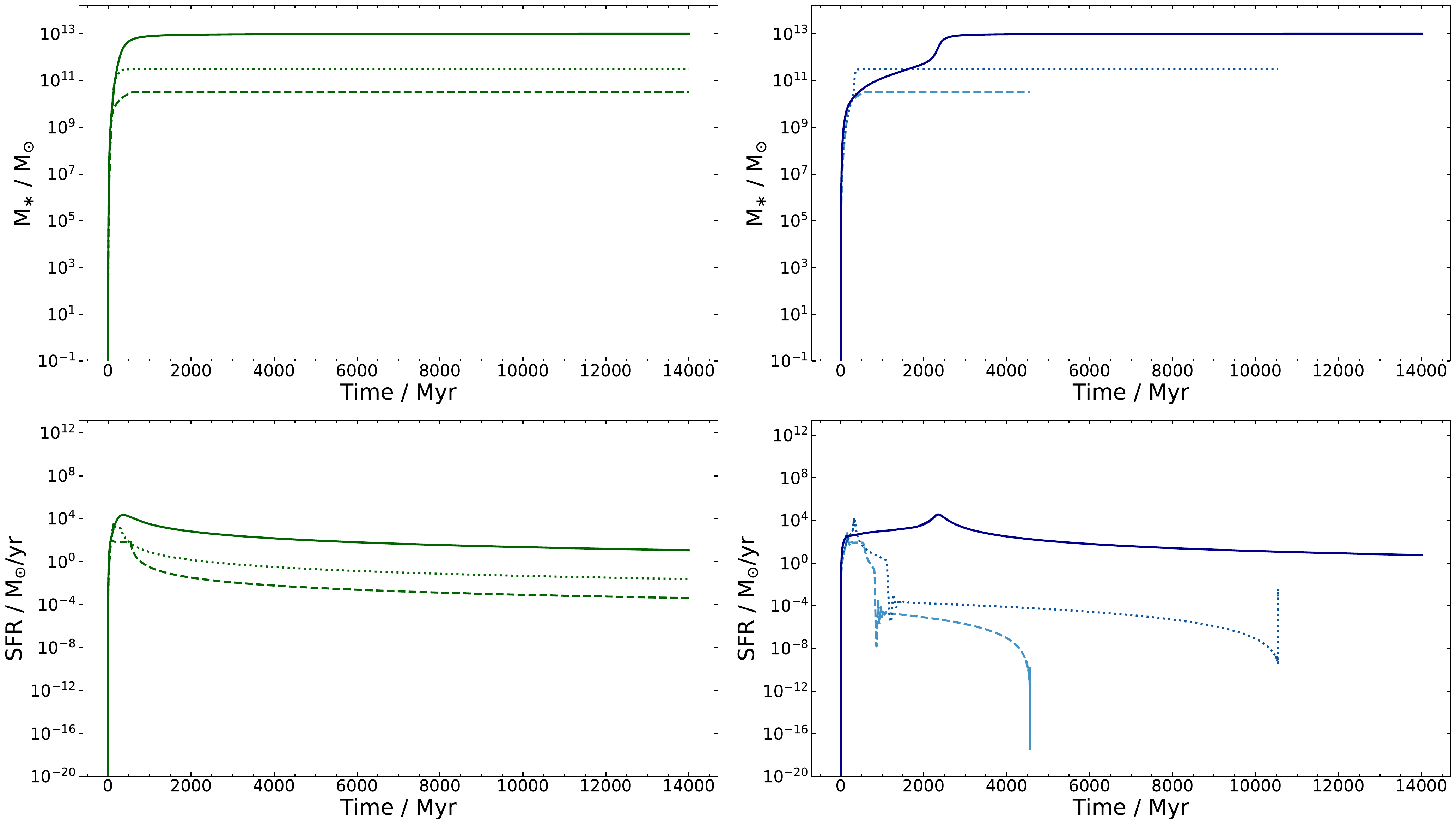}
    \caption{Similar plot to Fig.~\ref{fig:evol}, but for the three calculations with short and heavy accretion. The dashed line is the lightest galaxy, the dotted line the intermediate one and the solid line the heaviest galaxy. The final masses of the model galaxies are $10^{10.5}$, $10^{11.5}$ and $10^{13}$ $M_{\odot}$ for the canIMF models as well as for the IGIMF models. Despite the short duration of the gas infall, the gas returned from the evolving stars ensures continued and decreasing star-formation except in the two lowest accretion rate models which have a top-light gwIMF thus a lacking of massive stars. The continued star-formation in the other models implies that real elliptical galaxies must have a heating agent that keeps the gas temperature too high for star star-formation.}
    \label{fig:evol_ell}%
\end{figure}
As seen from Eq.~(\ref{eq:eqSFR}), the SFR reaches an equilibrium state determined fully by the accretion rate and the gwIMF assumption, but is independent of the local star-formation law.

\begin{figure}
   \includegraphics[width=\columnwidth]{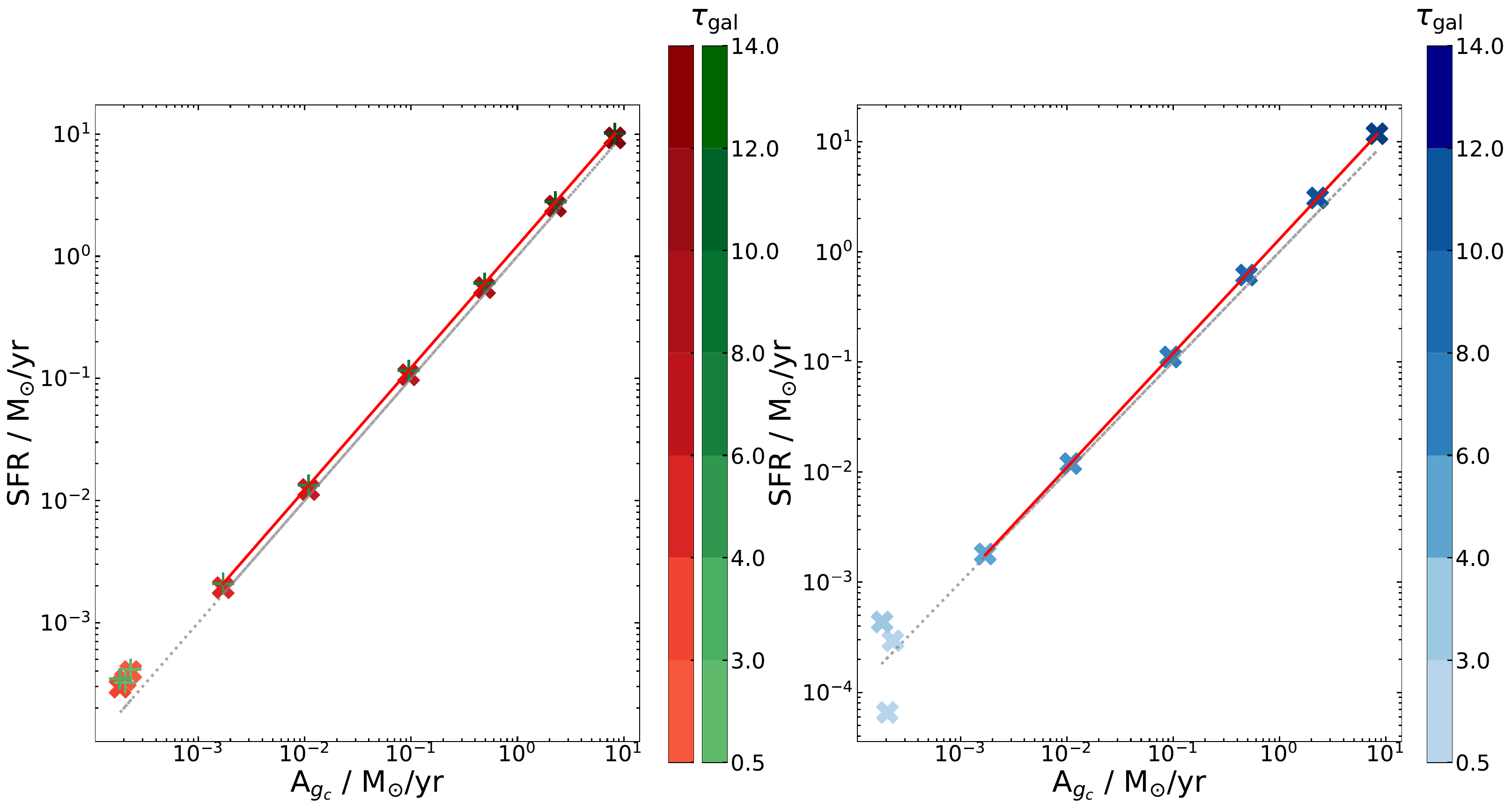}
    \caption{SFR as a function of accretion rate $A_{g_\mathrm{c}}$ for all models, for the two invariant IMFs in the left panel and for the IGIMF model in the right panel. The red lines are fits of Eq.~(\ref{eq:SFR_fit}) to the canonical IMF model and the IGIMF model, only considering galaxies that reached an equilibrium state. The gray dotted lines are the one-to-one relation and the colors are the same as in the previous figures.}
    \label{fig:Ac_SFR}%
\end{figure}

In Fig.~\ref{fig:Ac_SFR}, the equilibrium SFR is plotted as a function of accretion rate $A_{g_\mathrm{c}}$, as described in Eq.~\ref{eq:eqSFR}, using SFR instead of SFRD. The red lines are fits for all model galaxies that have reached an equilibrium state, using a power law relation
\begin{equation}\label{eq:SFR_fit}
    \rm SFR = \eta_1 \, A_{g_\mathrm{c}}^{\eta_2}.
\end{equation}
$\eta_1$ and $\eta_2$ are derived for the canonical IMF model and for the IGIMF model, with the values being $\eta_1 = 1.2191 \pm 0.0004$ and $\eta_2 = 0.9999 \pm 0.0001$ for the canonical IMF and
$\eta_1 = 1.3008 \pm 0.0118$ and $\eta_2 = 1.0362 \pm 0.0026$ for the IGIMF. This illustrates that the equilibrium SFR is determined entirely by the accretion rate as well as the influence of a top-heavy IMF in the IGIMF model.

The more top-heavy the gwIMF is, the larger becomes the deviation between accretion rate and equilibrium SFR, since the factor 1-$\eta$$\zeta$ decreases for a more top-heavy gwIMF. This leads not only to a higher equilibrium SFR, but also the time to reach the equilibrium becomes longer.\par
Another effect of the accretion driven self-regulation is the proportionality of the galaxy mass and the accretion rate. In Fig.~\ref{fig:M_Ac} the accretion rate is plotted logarithmically against the stellar-population-mass of the galaxy in the upper panels and against the total baryonic mass in the lower panels.

Independent of the choice of the gwIMF model, there is a power-law relation between the accretion rate and the stellar-population-mass as well as the total mass of the galaxy, hence the accretion rate is determining the mass of the galaxy. The solid gray lines represent the best fit between the logarithms of the accretion rate and of the stellar-population-mass, excluding the lightest three galaxies because these have not yet reached a self-regulated equilibrium (see Fig.~\ref{fig:evol}). The relation is:
\begin{equation}
    \rm log_{10}\left(\frac{A_{g_\mathrm{c}}}{\it M_{\odot}/ \rm Myr}\right) = 0.74 \cdot \rm log_{10}\left(\frac{M_{\star}}{\it M_{\odot}}\right) - 1.29
    \label{eq:canFIT}
\end{equation}

\noindent for the canIMF model and:

\begin{equation}
    \rm log_{10}\left(\frac{A_{g_\mathrm{c}}}{\it M_{\odot}/ \rm Myr}\right) = 0.72 \cdot \rm log_{10}\left(\frac{M_{\star}}{\it M_{\odot}}\right) - 1.09
    \label{eq:IGIMFFIT}
\end{equation}

\noindent for the IGIMF model. The proportionality constant between these two quantities is 0.74 $\pm$ 0.04 in the canIMF model and 0.72 $\pm$ 0.05 for the IGIMF model. The dotted lines show the relation between the accretion rate and the total mass, M$_{\rm tot}$ = M$_{\rm gas}$ + M$_{\star}$. In this case the proportionality constant between these two quantities is 0.79 $\pm$ 0.02 in the canIMF model and 0.79 $\pm$ 0.02 for the IGIMF model.\par
The red dashed lines in Fig.~\ref{fig:M_Ac} show a linear relation between $A_{g_\mathrm{c}}$ and the mass (either stellar-population-mass or total baryonic mass) with the proportionality constant being equal to the Hubble-Lema\^{i}tre constant $H_0$, the full relation being:

\begin{equation}
    \frac{A_{g_\mathrm{c}}}{\it M_{\odot}/ \rm Myr} = \frac{H_0}{\rm Myr^{-1}} \cdot \frac{M}{ M_{\odot}}.
    \label{eq:H0_fit}
\end{equation}

The value of $H_0 = (67.4 \pm 0.5)$ km s$^{-1}$ Mpc$^{-1}$ is taken from \cite{Planck}. The red dotted lines in Fig.~\ref{fig:M_Ac} show the same relation, but with a fitted $H_0$ value. For the relation between total baryonic mass and accretion rate, the fitted value is $H_0 = (123.045373 \pm 0.00001)$ km s$^{-1}$ Mpc$^{-1}$ for both canonical IMF and IGIMF. Self-regulated star-forming disk galaxies with M$_{\rm tot}$ > 3 $\cdot$ 10$^7$ $M_{\odot}$ thus need a specific accretion rate, ($A_{\rm c}/M_{\rm tot}$), that is comparable to the Hubble-Lema\^{i}tre constant to sustain their star-formation rates (see also Fig.~\ref{fig:evol} below).\par 
As mentioned already (see Sect.~\ref{sec:limits}), the calculations are carried out with a fixed radius. Assuming that galaxies are expanding over their lifetime and were smaller in the past would move our fitted $H_0$ closer to the Planck value.\par

The evolution in these self-regulated models is for all calculations similar, with the galaxies experiencing an epoch of high star-formation (higher than the equilibrium SFR) after the gas infall has started. Afterwards the system settles into an equilibrium state in SFR and metallicity enrichment. The evolution of the SFR (bottom panel) and of the stellar-population-mass (top panel) are shown in Fig.~\ref{fig:evol} for one heavy, one light and an intermediate galaxy with constant accretion, for the canIMF model in the left panel and the IGIMF model in the right panel. Due to a top-heavy gwIMF and the therefore higher gas return in the IGIMF cases the phase of enhanced star-formation is pronounced and the time for the system to reach equilibrium is prolonged. In contrast to \cite{accreg}, there is no SFR decline in the beginning caused by the dissipation of the accretion energy. This is because we start the accretion at t=0, and not after 500 Myr as in \cite{accreg}.\par

An important aspect of the self-regulated models studied here is that the constant SFRs in the self-regulated equilibrium phase evident in the lower panels of Fig.~\ref{fig:evol} agree with the constant SFRs of nearby galaxies (see footnote~\ref{footnote:3}) and that this requires the accretion rate, $A_{g_\mathrm{c}}$, for a galaxy to remain constant over cosmological time.\par

The same plot is shown in Fig.~\ref{fig:evol_ell} for the three model galaxies with a short and heavy accretion. These models relate to the formation of massive galaxies in the central regions of galaxy clusters where the proto-galactic gas clouds collapse but cannot continue accreting further gas after the collapse (c.f. \citealp{downsizing}, \citealp{Eappen25}). The  models also experience an epoch of enhanced star-formation, and in the canIMF model they settle in an equilibrium state with very little star-formation for the lighter two galaxies. Already in this model the heaviest galaxy doesn't reach a quenched state by self-regulation, which is suggesting the necessity of an additional heating source (possible heating sources could be AGNs, virial-shock heating or cosmic rays and other). The same is true for the heaviest galaxy in the IGIMF model.\par
For the heaviest IGIMF model galaxy with a short massive accretion, the blue solid line in Fig.~\ref{fig:evol_ell}, the evolution of the SFR shows a bump at 2000 Myr. This is due to the change in the gwIMF from top-heavy to top-light and bottom-heavy due to the high metallicity of the galaxy.\par
The lighter two galaxies in the IGIMF model shut down star-formation quickly and experience a short epoch of oscillating star-formation. Afterwards the SFR is too low to produce high-mass stars which would return gas to the system. Hence, these systems consume their gas depot, while the time to do so is shorter the smaller the galaxy is.\par
In these cases the calculations reach the limitations of the model by having no gas content in one of the gas phases and as a consequence the corresponding temperature cannot be calculated and the calculations are stopped.

\begin{figure}
   \includegraphics[width=\columnwidth]{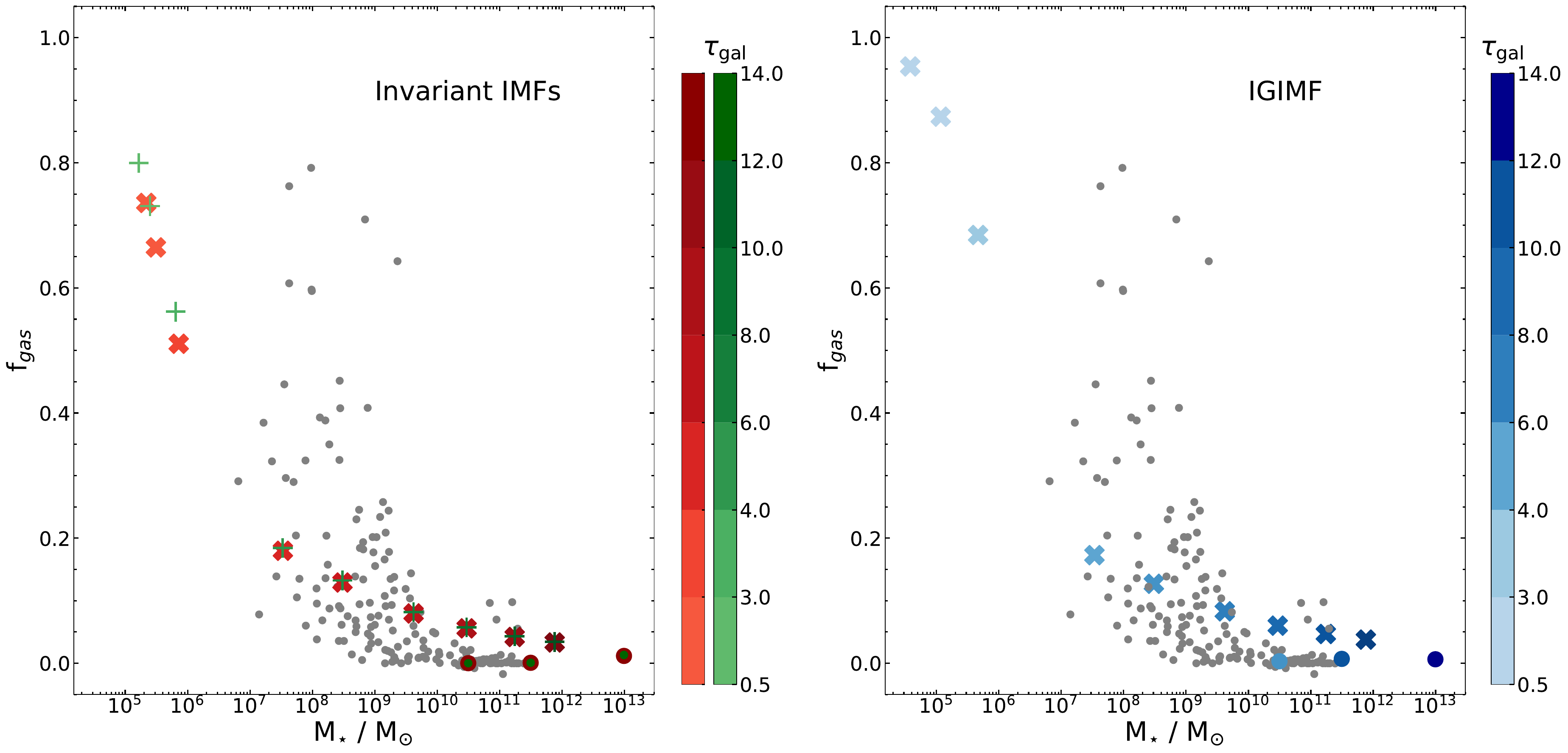}
    \caption{Gas fraction as a function of the stellar-population-mass. The gray dots are the gas fractions inside the scale radius for the SPARC galaxy sample (see \citealp{Lelli2016}), assuming $\Upsilon_{\star} = 0.5$ $M_{\odot}$/$L_{\odot}$. The other symbols and colors are corresponding to the plots in Fig.~\ref{fig:GMS14}, Fig.~\ref{fig:ZM14}. The younger/lighter the galaxies are, the higher is the gas fraction. Overall, the IGIMF calculations show higher gas fractions for the same initial parameters.}
    \label{fig:fgas}%
\end{figure}

\subsection{Gas fraction}\label{sec:fgas}
The gas fraction is defined as:

\begin{equation}\label{eq:fg}
    f_{\rm gas} = \frac{(c+g)}{(c+g+r+s)}.
\end{equation}

\noindent In Fig.~\ref{fig:fgas} the resulting gas fractions are compared to the local gas fractions at the scale radius of the SPARC galaxy sample (see \citealp{Lelli2016}), which the authors estimate 
as $f_{ \rm gas} = V_{\rm gas}^2 / V_{\rm bar}^2$, shown by the gray points.\par
As expected from observations, smaller galaxies show higher gas fractions, reaching over 90\,\% in the IGIMF model for the smallest/youngest galaxies. Overall, all three IMF cases reproduce the observed sample reasonably, leading to small gas-rich systems and gas-poor high-mass galaxies.

\begin{figure}
   \includegraphics[width=\columnwidth]{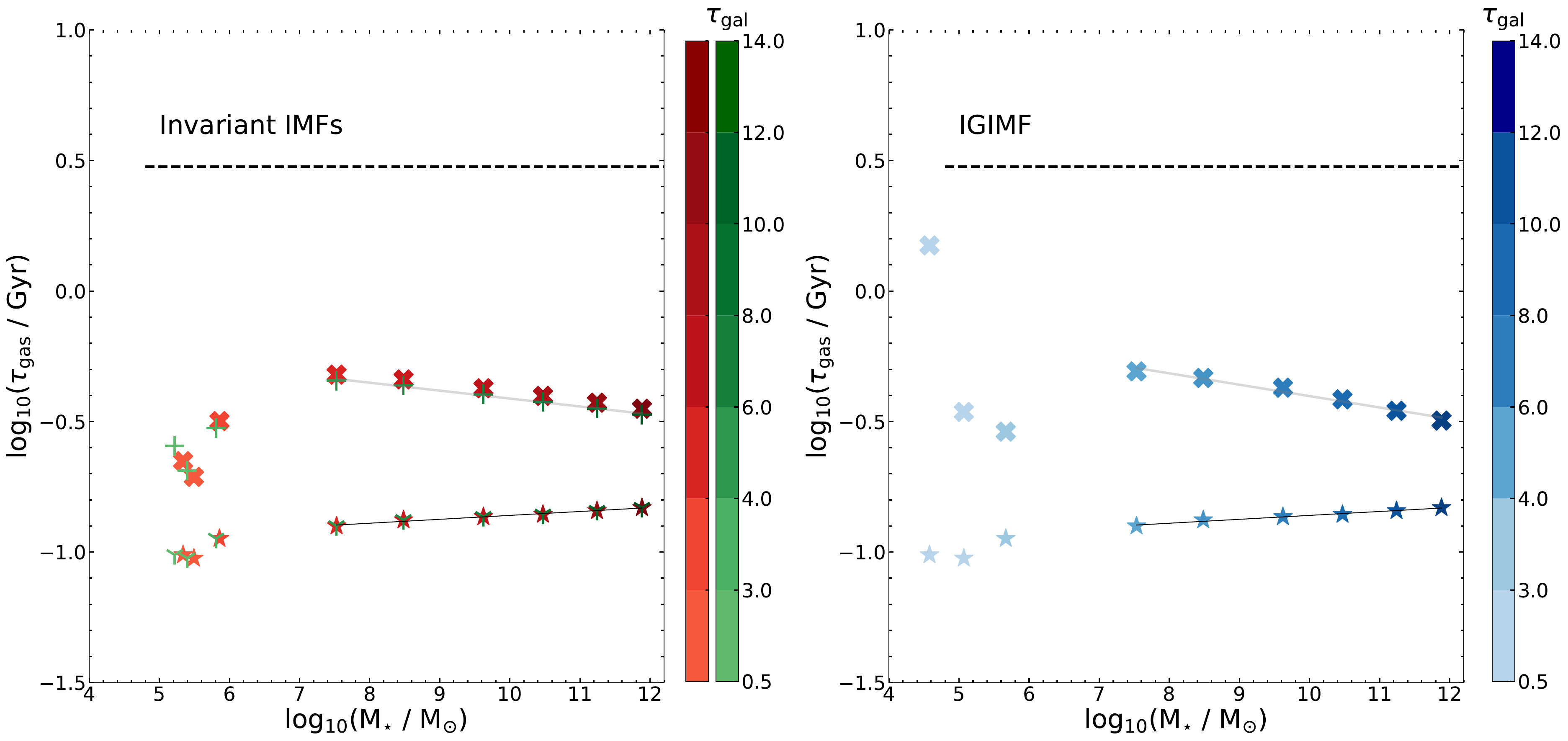}
    \caption{Gas depletion time $\tau_{\rm gas}$ is plotted as a function of the stellar-population-mass for all models with continuous infall according to the previous plots as x-symbols. The gray line is the best fit through the galaxies with a constant $\tau_{\rm gas}$. For the interpretation see Sect.~\ref{sec:taugas}. Compared to the gas depletion time, $\tau_{\rm gas}$, the star symbols represent the orbital period (see Eq.~(\ref{eq:T_orb})) of the same galaxy models and the black line is the corresponding fit. In both panels the black horizontal dashed line corresponds to a $\tau_{\rm gas}$ of 3 Gyr, as derived in \cite{JPA09}.}
    \label{fig:tau}%
\end{figure}
\subsection{Gas depletion time}\label{sec:taugas}

In Fig.~\ref{fig:tau}, the gas depletion timescale $\tau_{\rm gas}$, which is often referred to as the inverse of the star-formation efficiency, is defined as:

\begin{equation}\label{eq:tau_gas}
\tau_{\rm gas} = \frac{M_{\rm gas}}{SFR},
\end{equation}

\noindent is plotted for all constant accretion models, and compared for non-varying gwIMFs (left panel) and the IGIMF models (right panel). For the two non-varying gwIMF models the result is an almost constant gas depletion timescale of 779 $\pm$ 3 Myr, while for the IGIMF models, $\tau_{\rm gas}$ is 1080 $\pm$ 8 Myr. However, similar to Fig.~\ref{fig:size}, the smallest galaxies do not fit the relation, but have lower depletion timescales because they have not reached equilibrium during their short lifetime and are therefore excluded for the fitting.\par
The gas depletion timescales can be also expressed as a star-formation efficiency per 10 Myr, which is the time of one star-forming epoch, as:
\begin{equation}\label{eq:efficiency}
    \epsilon_{\rm 10 Myr} = \frac{10 \;\rm Myr}{\tau_{\rm gas}}.
\end{equation}

The values we obtain are for the canIMF model $\epsilon_{\rm 10 Myr}$ = 0.01283 $\pm$ 0.00005 and for the IGIMF model $\epsilon_{\rm 10 Myr}$ = 0.00926 $\pm$ 0.00007 $\approx 1\,\%$. This over a wide mass range almost constant efficiency suggests a self-regulating mechanism within disk galaxies to produce stars at a rate that is sufficient to stabilize the galaxy at an equilibrium state. This mechanism can be understood as an analogy to the interior of stars, where the nuclear fusion is stabilizing the star against gravitational collapse with exactly the right amount of released energy.\par

The black dashed line is at 3 Gyr, which is the derived depletion timescale from \cite{JPA09} for a galaxy to be on the MS.\par
Similar to their result, the depletion timescale shows an overall flat relation, implying that the star-formation efficiency is very similar for dwarf galaxies as well as for massive star-forming disk galaxies (also compare to \citealp{Hasl24}).\par
The model timescales are shorter than the above galaxy-wide estimates because here all parameters are derived solely within the effective radius. Real galaxies will have a much shorter $\tau_{\rm gas}$ within their effective radius and a higher $\tau_{\rm gas}$ in the outer regions, where more gas is present (see e.g., \citealp{Varytau}).\par
In comparison to the gas depletion time $\tau_{\rm gas}$ also the orbital period, which is defined as:

\begin{equation}\label{eq:T_orb}
    T = \frac{2\pi R}{V_{\rm c}},
\end{equation}

\noindent with the Baryonic Tully-Fischer Relation (\citealp{McGaugh00}, \citealp{Lelli19}),
\begin{equation}\label{eq:V_c}
    V_{\rm c} = (G(M_{\star} + M_{\rm gas})a_0)^{1/4}
\end{equation}

\noindent and a value of $a_0 = 1.2 \times 10^{-10} \;\rm m/s^2$ (\citealp{a0}), is also plotted as the star symbols in Fig.~\ref{fig:tau}. $a_0$ is the critical acceleration in modified Newtonian dynamics (MOND). As well as $\tau_{\rm gas}$, T is also almost constant over the mass range, approximately at 97 $\pm$ 3 Myr. That the gas-consumption timescales and orbital times at the effective radius are so constant over 7 orders of magnitude of stellar-population-mass with the former being about ten times the latter for the realistic (IGIMF) models constitutes an interesting result.

\section{Outflows}\label{sec:outflows}

\begin{figure}
   \includegraphics[width=\columnwidth]{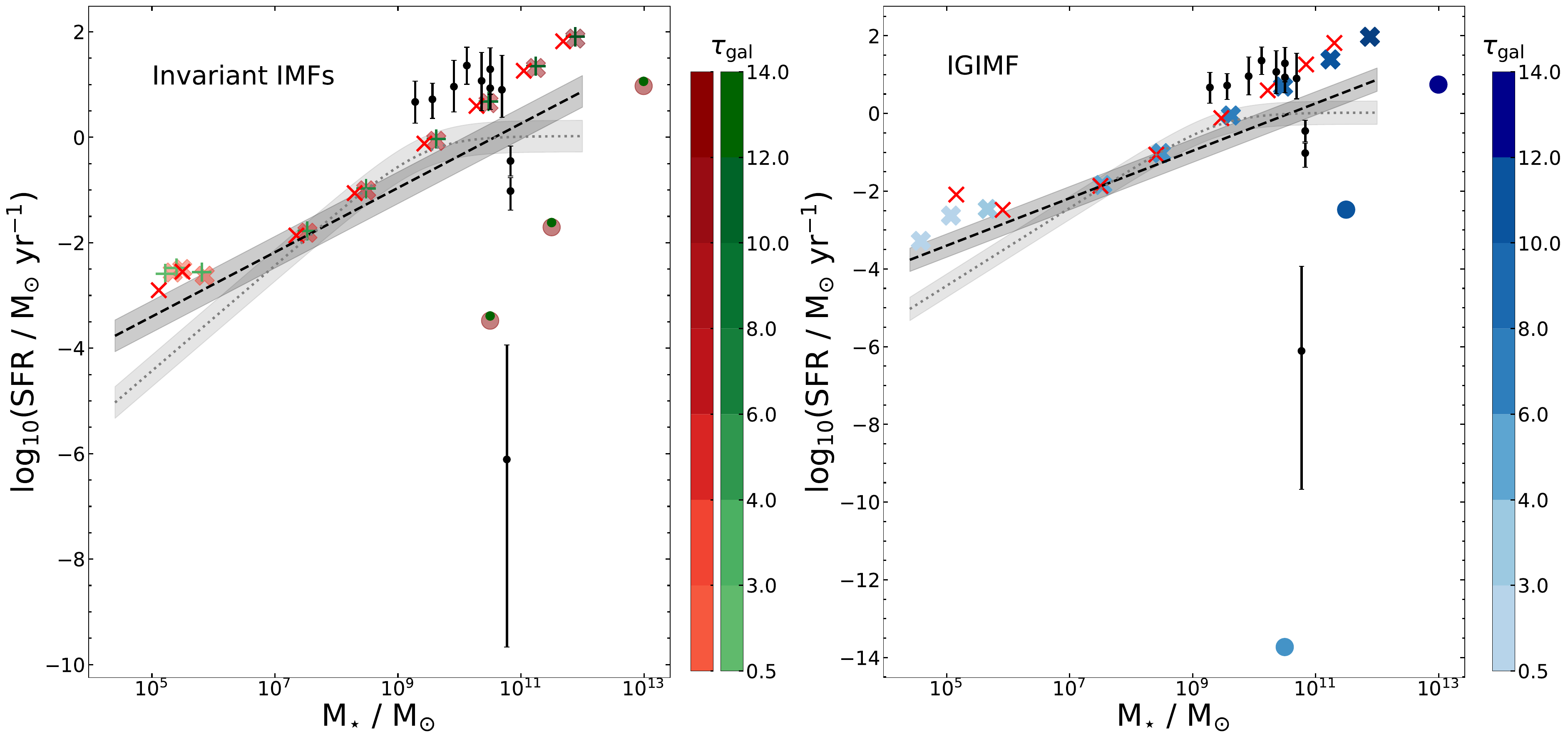}
    \caption{Same plot as in Fig.~\ref{fig:GMS14}, but in both panels the thin red crosses are the model galaxies recalculated with galactic outflows.}
    \label{fig:GMSoutlfow}%
\end{figure}

\begin{figure}
   \includegraphics[width=\columnwidth]{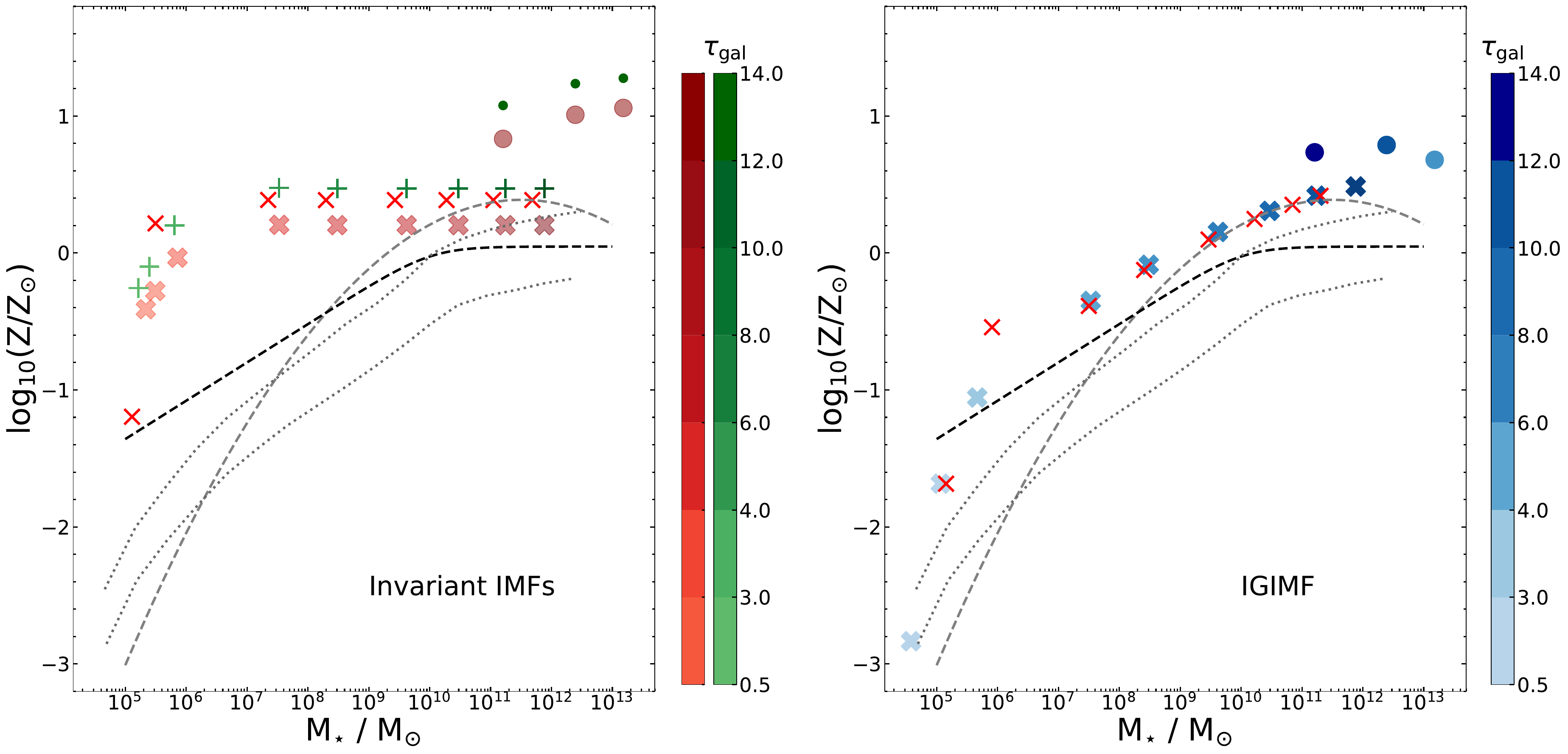}
    \caption{Same plot as in Fig.~\ref{fig:ZM14}, but in both panels the thin red crosses are the model galaxies recalculated with galactic outflows.}
    \label{fig:ZMoutflow}%
\end{figure}

To test the influence of outflows on the evolution of the model galaxies, the canIMF and IGIMF models with constant accretion are recalculated with the extra term $\eta_{\rm esc}$ acting on Eqs.~(\ref{eq:gdot}), (\ref{eq:rdot}), (\ref{eq:zg}) and (\ref{eq:zr}). The outflow in this scenario is caused by SN II explosions, causing some gas to escape the model galaxy instead of being returned to the gas phase $g_\mathrm{h}$. For this, the escape fraction of returned gas is calculated using Eq.~(\ref{eq:eta_esc}).

When the SN-driven hot gas expands with larger than escape velocity, it is considered as outflow/galactic wind and thus lost from the system. Since the expelled gas is metal-rich, this could lead to a decrease in overall gas metallicity (see \citealp{Hensler13}).\par

In Fig.~\ref{fig:GMSoutlfow} and Fig.~\ref{fig:ZMoutflow} similar plots to Fig.~\ref{fig:GMS14} and Fig.~\ref{fig:ZM14} are shown, but with recalculated models with outflows shown as the red crosses. For the MS in Fig.~\ref{fig:GMS14} there is almost no significant change between the models with and without outflows. As expected, the stellar-population-mass is slightly reduced due to the lower gas content for the model galaxies with a mass above $10^7$~$M_{\odot}$. Interestingly, for some of the lighter model galaxies, the stellar-population-mass is increased due to outflows. This should be a consequence of decreasing the gas temperature by expelling hot gas. The reduced gas content has no influence on these models, because of their young age, they do not have the time to form enough stars to reach an equilibrium state (see Fig.~\ref{fig:evol}).\par
For the MZR, in Fig.~\ref{fig:ZMoutflow} the models with outflow again are shown as the red crosses plotted over Fig.~\ref{fig:ZM14}. In the left panel for the canIMF the metallicity is decreased for the model galaxies above a mass of $10^7$ $M_{\odot}$.
For the lightest model galaxy in the canIMF model the metallicity is strongly decreased, as pointed out by \cite{Hensler13}. For the IGIMF model galaxies, the change in metallicity is not as enhanced as for the canIMF models. One exception on both the canIMF and IGIMF model is the second lightest model galaxy for which the metallicity is the same with and without outflows for the canIMF model and even higher for the IGIMF model. This has the same reason as the increased stellar-population-mass in this mass range, which leads to a stronger metallicity from enhanced star-formation. This effect is stronger in the IGIMF models due to the gwIMF being variable.\par
The inclusion of outflows slightly increases the agreement between the canIMF model and the observed MZR, leading to metal-poor small model galaxies. However, the shape of the modeled relation still differs from the observed relations in the mass range between $10^6$ and $10^{10}$ $M_{\odot}$, where the calculated metallicities remain too high.\par
The described outflow implementation is likely underestimating the outflow strength, since momentum-driven winds, fountains or radial gradients are not included and only SN II effects are taken into account.\par
However, the results obtained here are consistent with observational studies which have shown that outflows have an insignificant effect on the gas content and SFRs of star-forming and even star-bursting galaxies (see \citealp{dyndwarfout}, \citealp{gasdwarfout}, \citealp{kleverout}, \citealp{shakenout}).  This mitigates studies that claim outflows in dwarf galaxies are responsible for their low metallicity (see \citealp{MZTremonti04}, \citealp{Hensler13}, \citealp{McQuinn19}).

\section{Discussion}\label{sec:Dis}

The results presented in Sect. \ref{sec:Res} showed some fundamental differences between the models assuming a non-varying IMF and those based on the IGIMF.\par
One shortcoming of all models is the neglect of various dynamical energy distribution channels.
The full physical network would also include other processes, such as the formation of clouds by compression in supernova shells, dissipation by cloud-cloud collisions or stellar winds. These processes are not included here (see also Sect.~\ref{sec:limits}). However, \cite{coneva} showed that these processes do not essentially determine the type of the system’s behavior.\par
The models discussed in Sect.~\ref{sec:Res} also neglect outflows, for a discussion on this see Sect.~\ref{sec:outflows}.\par
Furthermore the influence of magnetic fields on the evolution of galaxies is under debate and not included in the model.\par
Another simplification is the neglect of intermediate and low-mass stars as contributors to the metal enrichment of the gas as well as a heating source, since only high-mass stars and SN II explosions are considered for the respective equations.\par
Real galaxies show radial variations in gas density, and consequently, the SFR and IMF are also expected to vary with radius. This spatial dependence is not yet accounted for in the current models. However, a fully consistent local formulation of the IGIMF theory, which could incorporate such radial variations, is yet to be developed (see \citealp{PflammAlt08}). Observed galaxies lie on the Radial Acceleration Relation (RAR, \citealp{Lelli17}), which connects their baryonic mass distribution to their dynamics. Including radial dependencies in future extensions of the model would represent a valuable improvement.

Overall, the IGIMF model reproduces the observational relations well, except for the two lowest mass galaxies. In this case they exhibit too large radii and an outlying $\tau_{\rm gas}$. This is due to their young age, which demands larger radii and a weaker star-formation maximum than expected from the other calculations in order to reach a sufficient mass with a low metallicity. At these small masses discrete effects begin to play a role such that the approach used here, which is based on the continuum assumption, breaks down.

\section{Conclusions}\label{sec:Con}
We seek to study if self-regulation galaxy evolution models can be constructed that capture the bulk observed properties of galaxies. To this end we calculate how the accretion of gas fuels star-formation that leads the gas to a self-regulation such that the accretion history defines the present-day properties of a galaxy. If one assumes that galaxy mergers are rare due to the absence of dark matter halos (\citealp{Kroupa15}, \citealp{Roshan21}, \citealp{KroupaGjergo23}, \citealp{Oehm24}, \citealp{Hernandez25}), the models here thus explore the scenario in which galaxies are fully self-regulated systems. The observed constant SFRs of nearby galaxies support this to be the case.\par
Another aspect is to investigate how the choice of the gwIMF influences the evolution of a galaxy in a chemical evolution model (\citealp{Yan21}). In our approach we compare an invariant canonical gwIMF with a variable gwIMF calculated using the IGIMF theory. In this context, it is worthwhile to mention that the gwIMF calculated using the canonical stellar IMF can include fractions of massive stars in a mass bin, while for the optimally sampled gwIMF calculated using the IGIMF theory this is not possible.\par
Regarding the MS of galaxies all models were able to reproduce a MS of star-forming galaxies similar to the one given in \cite{Hasl24}\par
For the calculation of elliptical type galaxies below a mass of $10^{12}$ $M_{\odot}$ all models reproduced a quenched 'elliptical' galaxy reasonably well. The calculations however required a short heavy infall of gas with no accretion afterwards. One scenario where this could be the case is for galaxies inside massive galaxy clusters, where in the inner parts the gas from the intracluster medium (ICM) can no longer reach the galaxies and therefore no significant later accretion is happening.\par
For the heaviest galaxy with a short and heavy accretion no quenching of star-formation was occurring during the lifetime of 14 Gyr. This suggests the presence of additional (hitherto unknown) heating sources in such systems, which would omit star-formation and lead to quiescent massive galaxies.\par
Regarding the MZR, the IGIMF model naturally reproduces the MZR by becoming top-light for low-mass galaxies and leading directly to a lower metallicity through a deficit of massive stars, as already suggested by \cite{Koeppen08}. For the canIMF and SalIMF models no such relation was reproduced, rather the metallicity in these calculations is almost independent of the mass of the galaxies. The inclusion of supernovae driven outflows in the calculations slightly, but not significantly, improved the result at the low-mass end for the canIMF model, leading to metal-poor light galaxies. For the IGIMF model and on the MS of galaxies the outflows had no significant impact (see Sect.~\ref{sec:outflows}).\par
The observed gas fractions inside the scale radius, estimated for the SPARC galaxies (\citealp{Lelli2016}) could be broadly reproduced by all three gwIMF models, leading to light gas-rich and heavy gas-poor systems.\par
The models reproduce the observed constant SFRs provided the accretion rates are constant over the age of the galaxy.\par
The gas depletion timescale, $\tau_{\rm gas}$, shows a very flat dependency on the stellar-population-mass over 7 orders of magnitude in galaxy mass for all gwIMF models, with only the lightest galaxies deviating. The found $\tau_{\rm gas}$ derived from linear regression are approximately 0.19 and 0.87 Gyr for the canIMF model and the IGIMF  model respectively. Accounting for the fact that the used model only accounts for the region inside the effective radius, $\tau_{\rm gas}$ is expected to be lower than 3 Gyr, found by \cite{JPA09}. Thus, within the effective radius $\tau_{\rm gas}$ is about ten orbital times for galaxies with stellar-population-masses spanning $10^4$ to $10^{12}$ $M_{\odot}$. A star-forming disk galaxy thus transforms about 1\,\% of its interstellar medium into a new population of stars every 10 Myr.\par
In order to sustain star-formation at the observed levels the gas accretion needs to be proportional to the galaxy's baryonic mass with the proportionality constant, or specific accretion rate, being comparable to the Hubble-Lema\^{i}tre value.

\begin{acknowledgements}
PK acknowledges support through grant 26-21774S from the Czech Grant Agency and also through the DAAD Eastern-European Exchange Scheme between Bonn and Prague. We thank Eda Gjergo for helpful discussions on the cooling function.
\end{acknowledgements}

%
%
\bibliographystyle{aa}
\bibliography{ref.bib}

\begin{appendix}

\section{Stellar yields}\label{sec:yield}
\begin{figure}
   \includegraphics[width=\columnwidth]{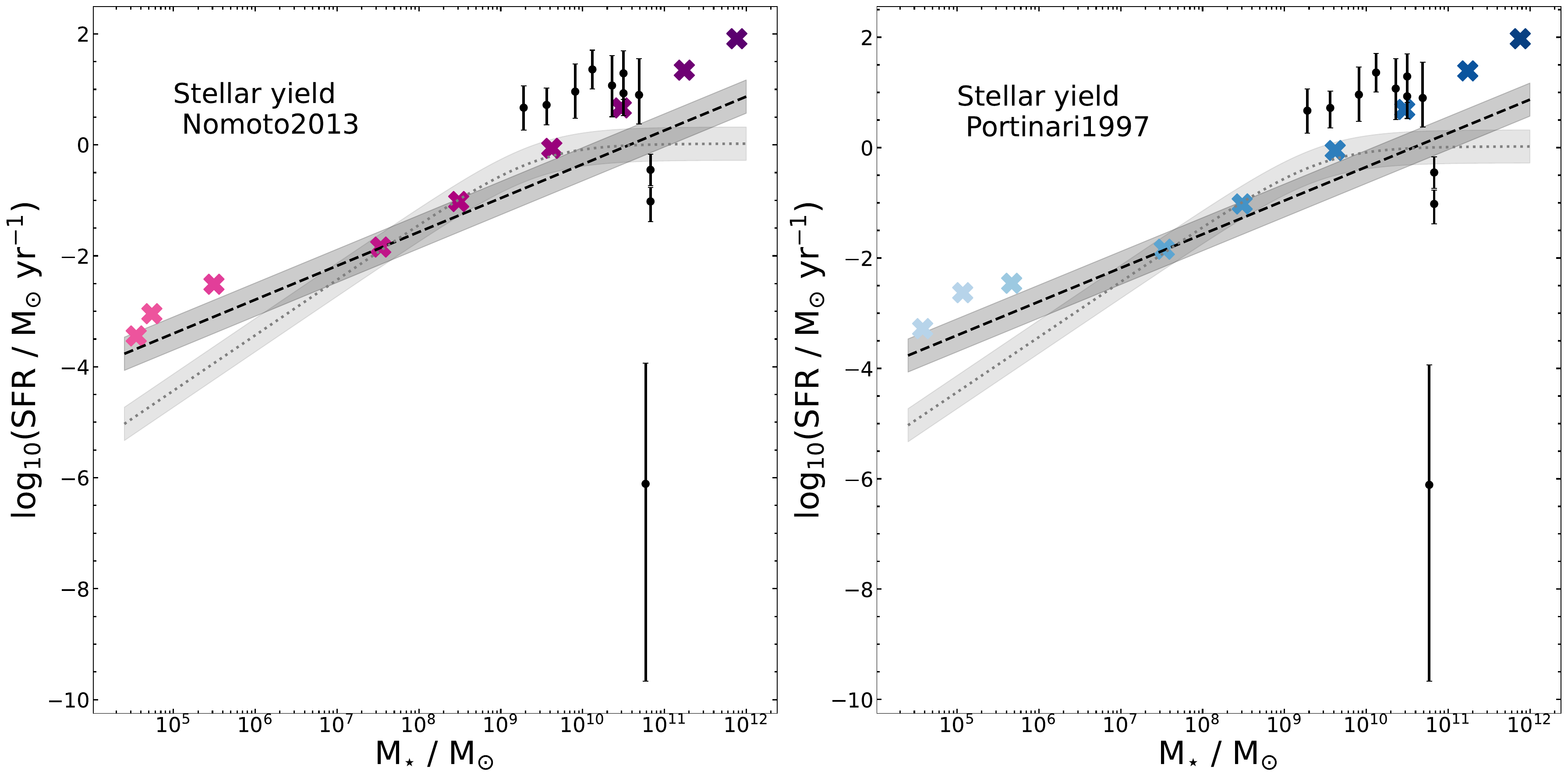}
    \caption{Stellar-population-mass-SFR relation for two different stellar yield tables (left panel from \cite{Nomoto2013}, right panel from \citealp{Portinari97}) and the IGIMF model. The stellar yields are used to compute the total yield for high-mass stars in the IGIMF model. The yields from \cite{Portinari97} are used for all other calculations. Lighter color shades correspond to younger ages.}
    \label{fig:GMSyield}%
\end{figure}

In Fig.~\ref{fig:yield} for both stellar yield tables (\citealp{Portinari97} and \citealp{Nomoto2013}) the computed total yields are plotted for different SFRs and for solar metallicity and compared to the canonical IMF and Salpeter IMF value.

\begin{figure}
   \includegraphics[width=\columnwidth]{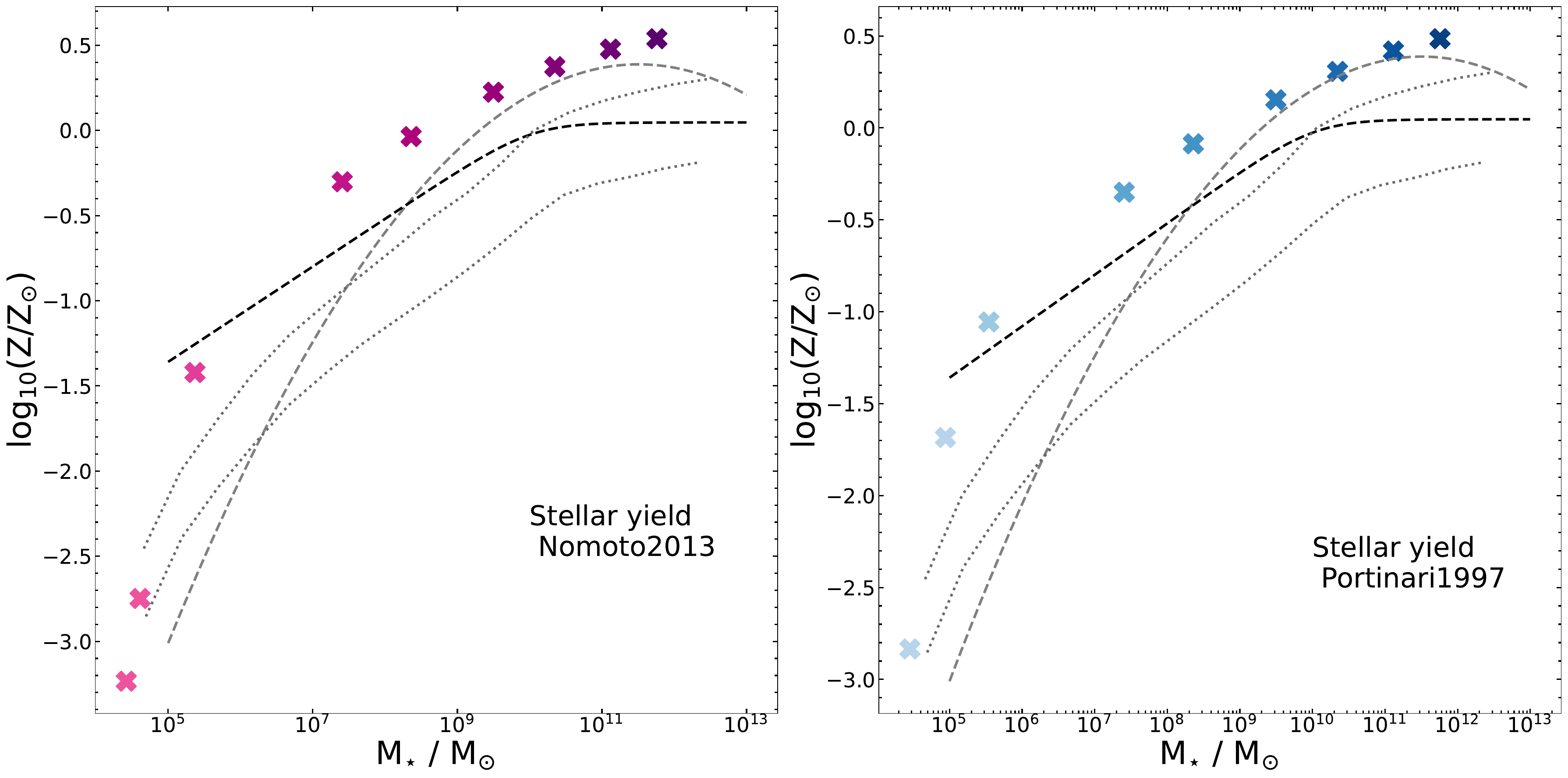}
    \caption{Similar plot as in Fig.~\ref{fig:GMSyield}, but this time again the MZR is plotted instead (compare to Fig.~\ref{fig:ZM14}).}
    \label{fig:ZMyield}%
\end{figure}

\begin{figure}
    \includegraphics[width=\columnwidth]{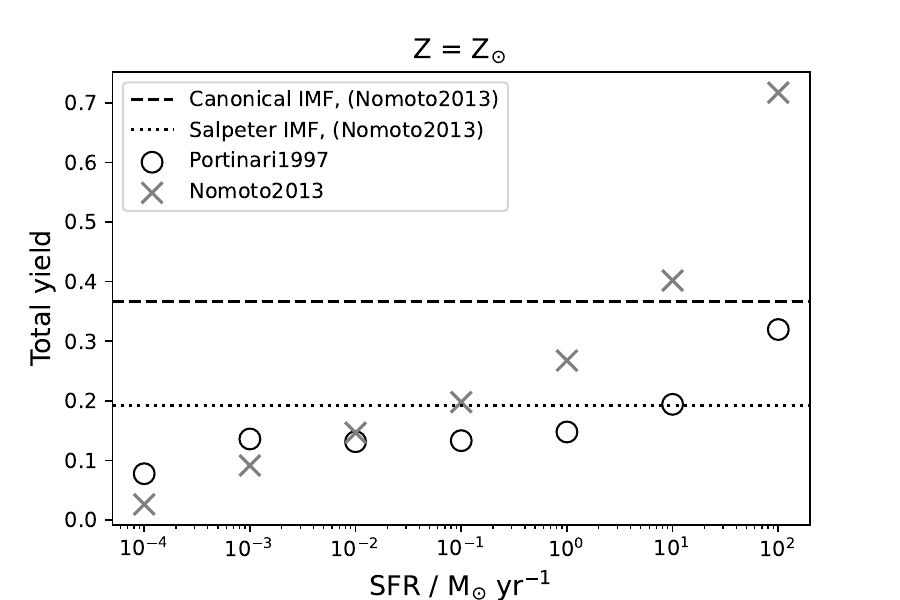}
    \caption{Total yield, $y$, of stars with masses larger than 10 $M_{\odot}$, computed as a function of SFR and for solar metallicity. For the calculations the IGIMF is assumed. The circles are calculated with stellar yields from \cite{Portinari97} and the crosses with the ones form \cite{Nomoto2013}. The dotted and dashed lines show, respectively, the total yield for the invariant Salpeter and the canonical gwIMFs.}
    \label{fig:yield}
\end{figure}

In Fig.~\ref{fig:GMSyield} and Fig.~\ref{fig:ZMyield}, the SFR and gas metallicity are plotted against the stellar-population-mass, but this time for two different stellar yield tables that were used to compute the total yield for the IGIMF model. For the left panel the stellar yields from \cite{Nomoto2013} were used while for the right panel the stellar yields from \cite{Portinari97} were adopted, the same as for the calculations in Sect. \ref{sec:GMS} -- \ref{sec:self_regualtion}.\par

Since the difference between both stellar yield tables is only minor, it is concluded that there is no significant impact on the results by the choice of the exact stellar yields, which is adding a robustness to the model because the exact description of the stellar yields is still under debate (\citealp{chemevo}).

\section{Different infall velocities}\label{sec:vac}
\begin{figure}
   \includegraphics[width=\columnwidth]{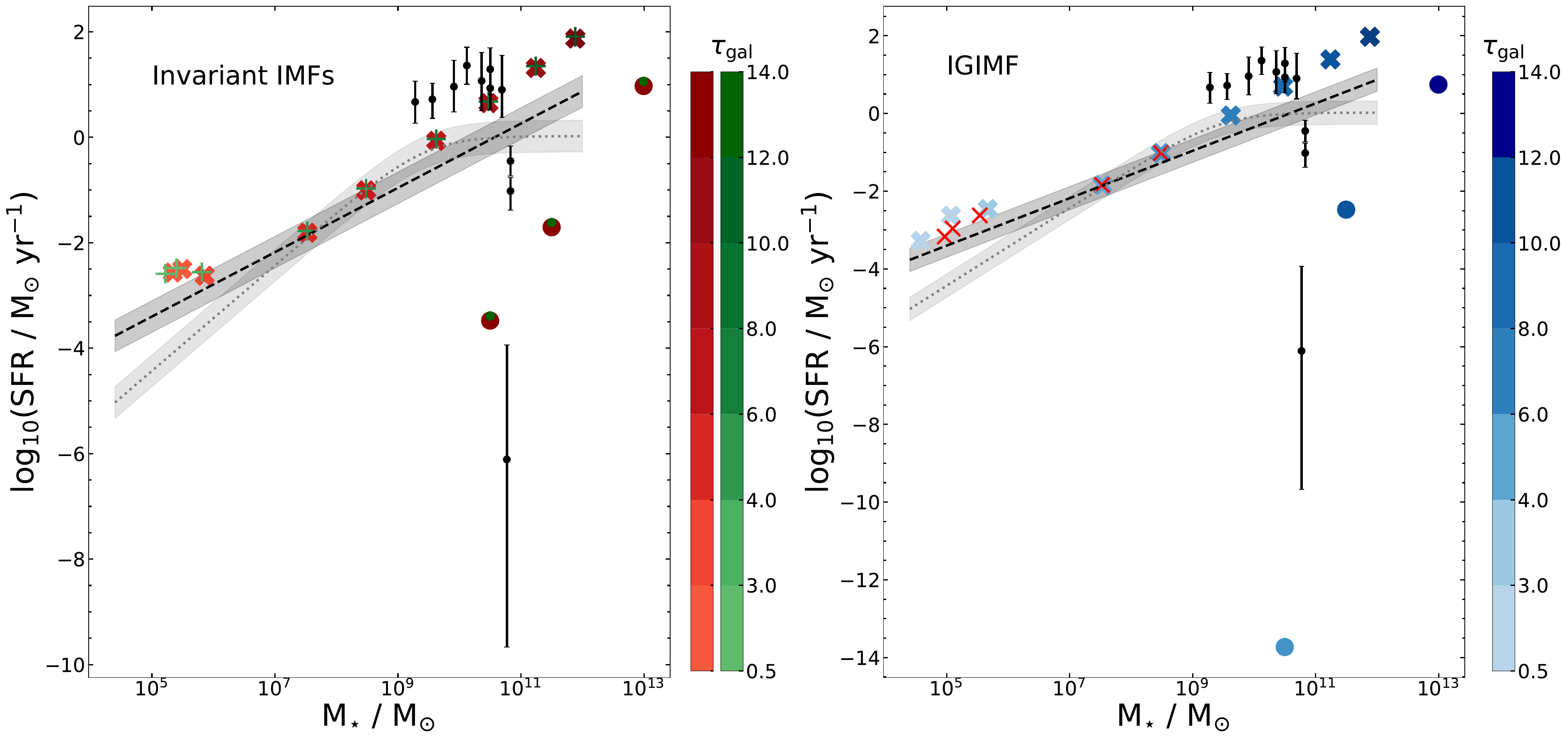}
    \caption{Same plot as in Fig.~\ref{fig:GMS14}, but in the right panel the red crosses are the smallest galaxies recalculated with a infall velocity of 10 km/s instead of 100 km/s (see Table \ref{tab:inicon}). The two heaviest recalculated galaxies yield the same result as before, therefore the red crosses are behind the blue crosses and are not visible.}.
    \label{fig:GMSvac}%
\end{figure}

\begin{figure}
   \includegraphics[width=\columnwidth]{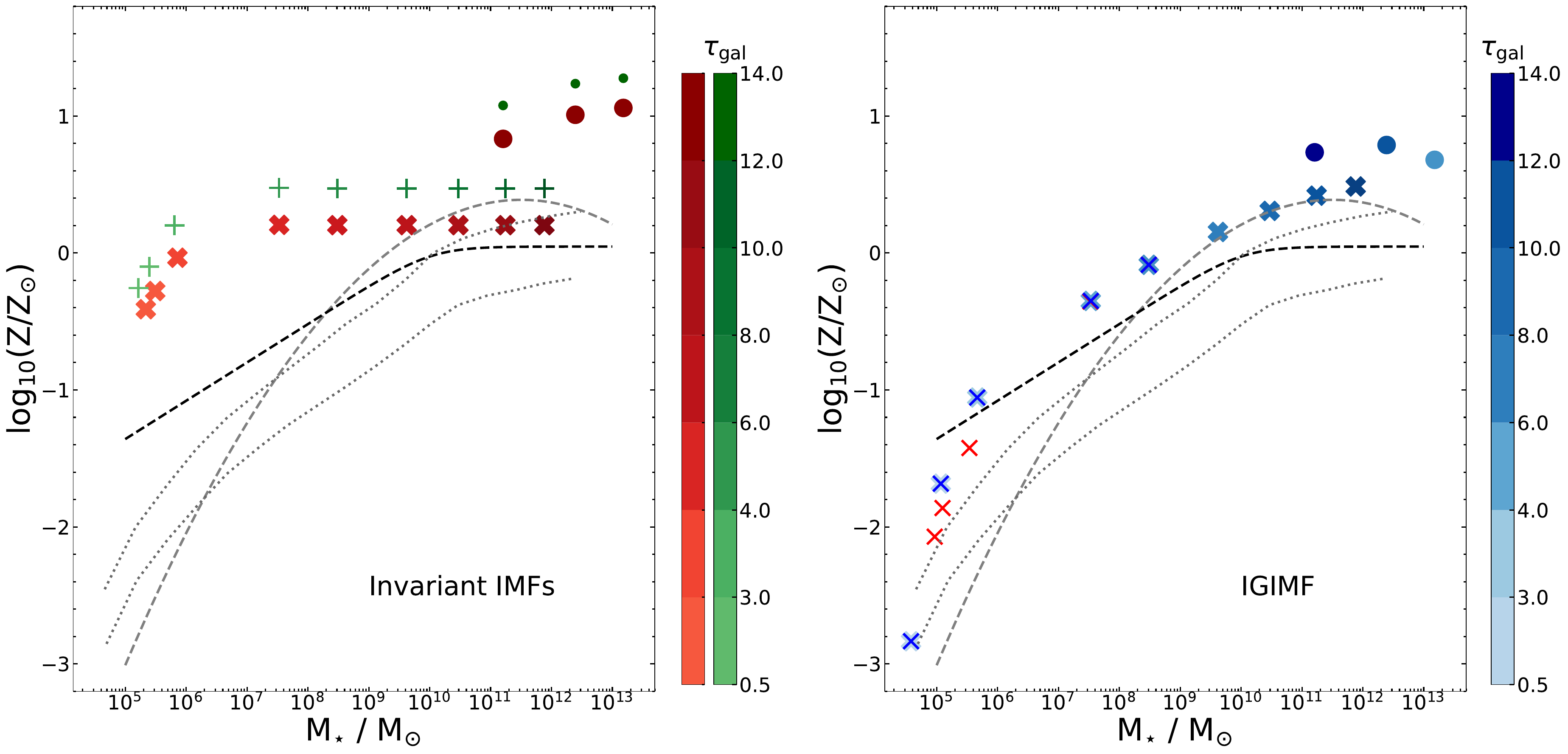}
    \caption{Same plot as in Fig.~\ref{fig:ZM14}, but in the right panel the red crosses are the smallest galaxies recalculated with an infall velocity of 10 km/s instead of 100 km/s (see Table \ref{tab:inicon}). The two heaviest recalculated galaxies yield the same result than before, therefore the red crosses are behind the blue crosses and are not visible.}
    \label{fig:ZMvac}%
\end{figure}

For the lightest five galaxies in the IGIMF model the calculations are repeated with a smaller gas infall velocity, namely $v_{\rm ac} =$ 10 km/s instead of 100 km/s, since the infall velocity should in principle depend on the mass of the galaxy. To investigate how the results change, these additional calculations are plotted on top of the previous results in Fig.~\ref{fig:GMSvac} in comparison to Fig.~\ref{fig:GMS14} and in Fig.~\ref{fig:ZMvac} in comparison to Fig.~\ref{fig:ZM14}.

As one can see in both Fig.~\ref{fig:GMSvac} and Fig.~\ref{fig:ZMvac}, the change in the infall velocity has a small influence on the individual galaxies and does not affect the overall shape of the relations. For the heaviest two recalculated galaxies there is no difference at all. Therefore it should not undermine the results to use a constant infall velocity of 100 km/s for all calculations.

\section{Initial parameters}\label{sec:iniparam}

In Table~\ref{tab:IGIMFparam}, Table~\ref{tab:inicon} and Table~\ref{tab:input}, the parameters for the galIMF code as well as the initial conditions for the calculations are listed.

\begin{table}
	\centering
	\caption{Intrinsic parameters to compute the galaxy-wide IMF.}
	\label{tab:IGIMFparam}
	\begin{tabular}{lccccccr}
		\hline
		alpha3\_model & delta\_t & I\_ecl & M\_ecl\_U & M\_ecl\_L & beta\_model & I\_str & M\_str\_L \\
		\hline
		2 & 10 & 1 & $10^9$ & 5 & 1 & 1 & 0.08 \\
		\hline
        \hline
        alpha\_1 & alpha1\_model & M\_turn & alpha\_2 & alpha2\_model & M\_turn2 & M\_str\_U &  \\
        \hline
        1.3 & 'Z' & 0.5 & 2.3 & 'Z' & 1 & 150 & \\
        \hline
	\end{tabular}
\tablefoot{The version of galimf we used is 1.1.10 and available on \href{https://github.com/Azeret/galIMF}{https://github.com/Azeret/galIMF}.}
\end{table}

\begin{table}
	\centering
	\caption{Initial parameters of the calculations.}
	\label{tab:inicon}
	\begin{tabular}{lccccccr}
		\hline
		$g_\mathrm{c}$ & $g_\mathrm{h}$ & $s$ & $r$ & $T_{g_\mathrm{c}}$ & $T_{g_\mathrm{h}}$ & $E_{g_\mathrm{c}}$ & $K_{g_\mathrm{h}}$ \\ 
		\hline
		7.6 $\times$ 10$^{-6}$  $\frac{M_{\odot}}{\rm pc^3}$ & 7.6 $\times$ 10$^{-7}$  $\frac{M_{\odot}}{\rm pc^3}$ & 0 & 0 & 3000 K& $10^7$ K  & 0 & 0 \\ 
        \hline
        \hline
        $T_{\rm ac}$ & $v_{\rm ac}$ & $t_{\rm infall}$ & $Z_{g_\mathrm{c}}$ & $Z_{g_\mathrm{h}}$ & $Z_s$ & $Z_r$ \\
        \hline
        3000 K & $100 \frac{\rm km}{\rm s}$ & 0 Myr & $10^{-7}$ & $10^{-7}$ & $10^{-7}$ &  $10^{-7}$\\
		\hline
	\end{tabular}
\tablefoot{The parameters were the same for all disk galaxy models. T$_{\rm ac}$ is the temperature of the accreted gas, v$_{\rm ac}$ the infall velocity and t$_{\rm infall}$ the time the accretion starts in the calculations, in all models from the beginning.}
\end{table}

\begin{table}
\caption{Input parameters for the calculations that are done for all three models (IGIMF, canIMF, SalIMF).}
\centering
\begin{tabular}{c|c|c|c|c}
\hline
$R$ / pc & $A_{g_\mathrm{c}}$ / $M_{\odot}$ Myr$^{-1}$ pc$^{-3}$& $\tau_{\rm gal}$ / Gyr& $\tau_{\rm ac}$ / Gyr & $g_\mathrm{c}$+$g_\mathrm{h}$  / $M_{\odot}$ pc$^{-3}$\\
\hline
\hline
43835  & 2.21 $\times$ 10$^{-4}$  & 14 & 0.54 & 3.74 $\times$ 10$^{-5}$\\
\hline
5595 & 1.42 $\times$ 10$^{-3}$  & 14  & 0.303 &2.09 $\times$ 10$^{-6}$\\
\hline
1418 & 4.89 $\times$ 10$^{-3}$  & 14  & 0.128 &3.08 $\times$ 10$^{-7}$\\
\hline
\hline
8071 & 3 $\times$ 10$^{-5}$  & 12 & 12 &8.36 $\times$ 10$^{-6}$\\
\hline
5445 & 2.7 $\times$ 10$^{-5}$  & 10 & 10 &8.36 $\times$ 10$^{-6}$\\
\hline
3393 & 2.4 $\times$ 10$^{-5}$& 8 & 8 &8.36 $\times$ 10$^{-6}$\\
\hline
2051 & 2.1 $\times$ 10$^{-5}$ & 6 & 6 &8.36 $\times$ 10$^{-6}$\\
\hline
1050& 1.8 $\times$ 10$^{-5}$ & 4 & 4 &8.36 $\times$ 10$^{-6}$\\
\hline
583 & 1.65 $\times$ 10$^{-5}$ & 3 & 3 &8.36 $\times$ 10$^{-6}$\\
\hline
227 & 3 $\times$ 10$^{-5}$ & 1 & 1 &8.36 $\times$ 10$^{-6}$\\
\hline
170 & 8 $\times$ 10$^{-5}$ & 0.5 & 0.5 &8.36 $\times$ 10$^{-6}$\\
\hline
170 & 9 $\times$ 10$^{-5}$ & 0.5 & 0.5 & 8.36 $\times$ 10$^{-6}$\\
\hline
\end{tabular}
\tablefoot{The first three rows are for the elliptical-like galaxies. Note that while the input time $\tau_{\rm gal}$ was always 14 Gyr, the calculations for the lighter two elliptical-like galaxies in the IGIMF model stopped earlier, at around 4 Gyr for the lightest one and at 10 Gyr for the intermediate one (see Sect.~\ref{sec:self_regualtion}). The last column gives the total initial gas density, $g_\mathrm{c} + g_\mathrm{h}$, for which $g_\mathrm{c}$ was set always one magnitude larger than $g_\mathrm{h}$ (see also Table~\ref{tab:inicon}).}
\label{tab:input}
\end{table}

\end{appendix}
\end{document}